\documentclass[reprint,amsmath,amssymb,aps]{revtex4-2}

\usepackage{graphicx}
\usepackage{bm}
\usepackage{hyperref}
\usepackage{amsmath}
\usepackage{braket}
\usepackage{dsfont}
\usepackage{siunitx}

\newcommand{\ee}{\mathrm{e}}
\newcommand{\ii}{\mathrm{i}}
\def\twobytwo#1#2#3#4{\left( \begin{array}{cc} #1 & #2 \\ #3 & #4 \end{array}\right)}

\begin{document}

	\title{Topological edge-state contributions to high-order harmonic generation in finite flakes}

	\author{Christoph J\"ur\ss}
	\affiliation{Institute of Physics, University of Rostock, 18051 Rostock, Germany }
	\author{Dieter Bauer}
	\affiliation{Institute of Physics, University of Rostock, 18051 Rostock, Germany }
	\date{\today}
	
	\begin{abstract}
        Edge states play a major role in the electron dynamics of topological insulators as they are the only conducting part in such materials. In this work, we consider the Haldane model for a 2D topological insulator, subjected to an intense laser field. We compare the numerically simulated high-harmonic generation (HHG) in the bulk of the Haldane model to HHG in corresponding finite flakes with edge states present, and explain the differences. In particular, peaks for energies below the bulk band gap appear in the spectra for the finite flakes. The positions of these peaks show a strong dependence on the size of the flakes, which can be explained using the  dispersion relation for the edge states.
	\end{abstract}

	\maketitle

	\section{Introduction} 
	
	In recent years, high-order harmonic generation (HHG) in condensed matter attracted more and more interest in the ultrafast, strong field physics community. The seminal paper reporting the observation of HHG in a bulk crystal appeared in 2011 \cite{Ghimire2011}. The process of HHG can be used to image the static properties \cite{VampaPhysRevLett.115.193603,TancPhysRevLett.118.087403,Lakhotia2020} and dynamical processes \cite{SchubertO.2014,Hohenleutner2015,You2017,Baudisch2018} inside condensed matter and opens paths towards new light sources \cite{Luu2015,Ndabashimiye2016,LangerF.2017}. For the generation of high harmonics, the target is illuminated by an intense laser pulse. In a solid, the laser excites electrons from the valence into the conduction band. The hole in the valence band and the electron in the conduction band propagate and eventually recombine upon the emission of a photon (interband harmonics). This simple three-step model for high-order harmonic generation in solids \cite{vampa_merge_2017}  is similar to the one in the gas phase \cite{corkum_plasma_1993,LewensteinPhysRevA.49.2117}. It was shown though \cite{Gaarde_imperfect_recol} that the collision of the electron with the hole can be imperfect and still generate harmonic radiation. Light can also be emitted by the movement of the electron and hole in bands, so called intraband harmonics, and by the time-dependent populations and transitions \cite{Jurgens2020}. The theory of HHG in solids is summarized in the tutorial \cite{yue2021introduction}. The review \cite{Goulielmakis2022} also focuses on the experimental progress in this field.
	
	Often, HHG is discussed in terms of the bulk properties of the system. However, the edges may have a significant influence on the emission of light as well \cite{bauer_high-harmonic_2018,DrueekeBauer2019,JuerssBauer2019,Bai2021,Schmid2021,Juerss_edge_nanoribbon}. In particular, topological insulators might be of interest, which are insulating in their interior (i.e., their bulk) but allow for current flow along the edges (in 2D materials) or surfaces (in 3D). These edge currents are topologically protected against various kinds of perturbations \cite{topinsRevModPhys.82.3045}. 
	
	The paradigmatic example for a topological insulator in 2D is the Haldane model \cite{Haldane_1988}, which is formulated in tight-binding approximation and was originally introduced as a toy model for a topological insulator without external magnetic field. Later it turned out that the hoppings Haldane introduced in his model are similarly realized in real materials through spin-orbit coupling \cite{QSHE_Graphene}. Moreover, the Haldane model can be experimentally realized using cold atoms \cite{Jotzu2014} or photonic platforms \cite{Rechtsman2013}. In the following, we refer to this hypothetical material obeying the Haldane model as Haldanite.  Haldanite has a honeycomb structure like graphene. Finite honeycomb lattices may have    zigzag or armchair edges both allowing for   edge states \cite{Hao2008,Yao_edgeStatesGraphene}. In this work, we investigate a finite flake with the shape of a hexagon, as shown in Fig. \ref{fig1}. All edges of that flake are of zigzag type.

	In order to transform topologically trivial graphene into topologically non-trivial Haldanite, an alternating on-site potential is added to break the sublattice symmetry and to open a band gap. The model is now describing materials like hexagonal boron nitride, which have a honeycomb structure too but consist of two different elements. Yet, in order to render the system topological, a complex hopping between next-nearest neighbors is introduced that breaks time-reversal symmetry. 
	
	Harmonic generation in Haldanite has been studied for the bulk \cite{Silva2019, chacon_observing_2018} and finite ribbons \cite{juerss2020helicity, Juerss_edge_nanoribbon}. Further studies were devoted to HHG in three-dimensional topological insulators. In an experiment on HHG in Bi$_2$Te$_3$, harmonics attributed to topological surface states were found to depend on the carrier envelope phase, leading to non-integer harmonics \cite{Schmid2021}. In Bi$_2$Se$_3$ \cite{Ghimire2021_PRA,Ghimire2021_nano} and BiSbTeSe$_2$ \cite{Bai2021}, the dependence of the surface and bulk harmonic spectra on the polarization and orientation of the laser pulse was investigated. In our current work, the focus lies on finite Haldanite flakes and a comparison of the HHG spectra with those from the bulk. Edge states exist in the finite flakes considered.  Energetically, the edge states lie in the band gap between the valence and the conduction band of the bulk and are expected to strongly modify HHG as compared to HHG from the bulk alone. 
	
	Indeed, our results show additional peaks for the finite flakes that are absent in the spectrum of the bulk system. In particular, two peaks are observed that shift in energy as the size of the system is changed. A direct connection between the edge states and those peaks is found.
	
	The manuscript is structured as follows: in Sec. \ref{sec:methods}, the methods to describe the system theoretically are presented. The results for the static system are discussed in Sec. \ref{sec:static_flakes} before the emission spectra are analyzed in Sec. \ref{sec:spectra}. First, the spectra for the finite flakes are compared to the spectrum for the bulk in Sec. \ref{sec:comp_spectra}. Second, the dependence of the spectra on the flake size is investigated in Sec. \ref{sec:spectra_size}, and third, the main features in the spectra are analyzed in Sec. \ref{sec:filtered_spectra} by filtering them out and investigating the corresponding currents that lead to these features. The main result on the currents along the edges that lead to the observed spectral features for different system sizes is discussed in Sec. \ref{sec:connection} before the paper is summarized in Sec. \ref{sec:summary}. Atomic units (a.u.; $\hbar=|e|=m_e=4\pi\epsilon_0=1$) are used throughout this paper unless stated otherwise.
	
	\section{Methods}\label{sec:methods}
	
	    The theoretical description of the system is similar to our previous works \cite{juerss2020helicity,juerss_bauer_ArmZZ,Juerss_edge_nanoribbon}. In particular, the method to calculate high-harmonic spectra is the same.
	
	    \subsection{Finite system without external field}

	    \begin{figure}
        \includegraphics[width = 0.8\linewidth]{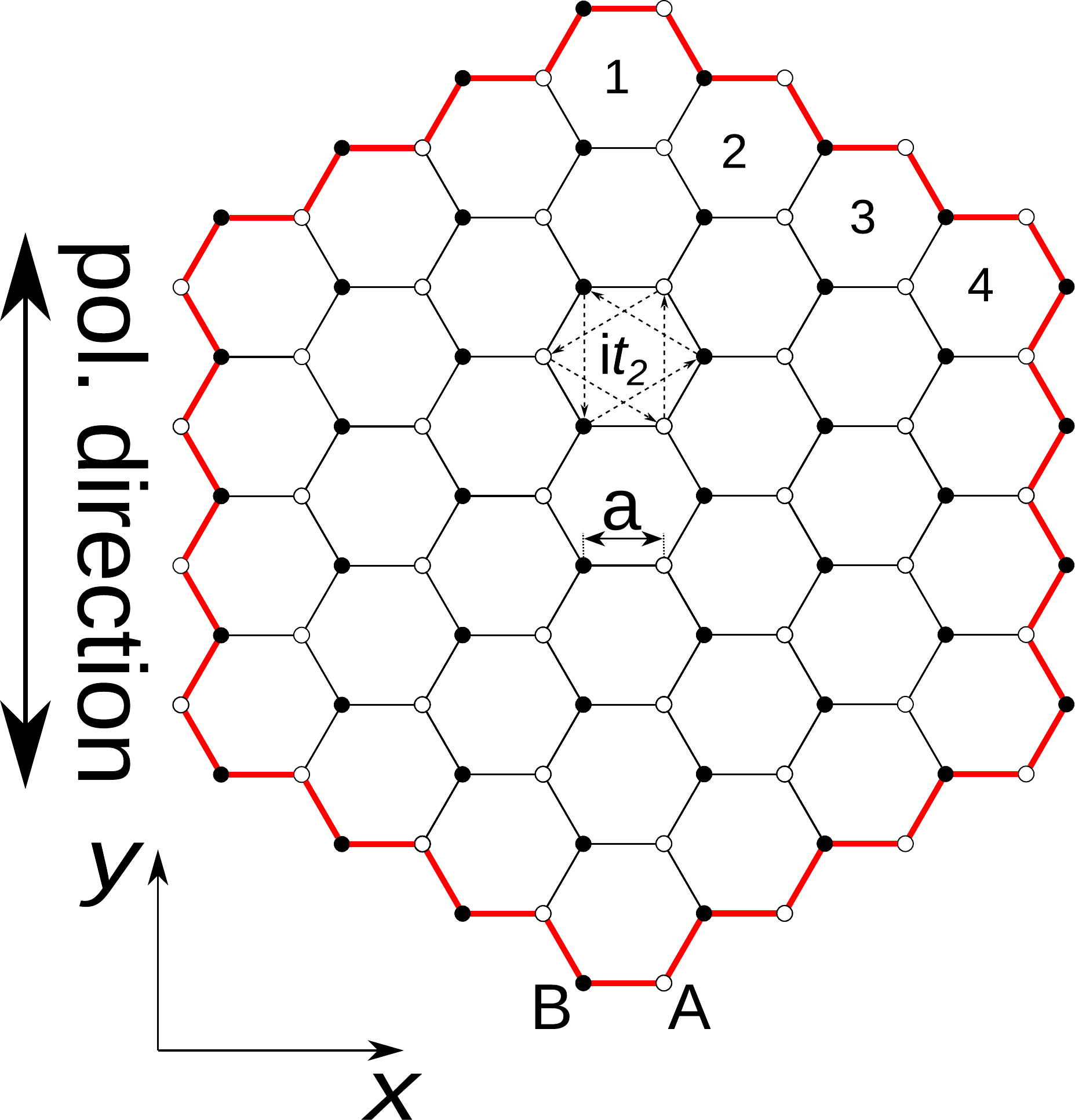}
        \caption{\label{fig1} Sketch of a finite Haldanite flake of size $N = 4$ (number of hexagons on each edge). There are two sublattice sites: A (open circles) and B (filled circles). The hopping between nearest neighbors (amplitude $t_1\in\mathds{R}$) is indicated by solid lines connecting the sites, the distance between two sites being $a$. The next-nearest neighbor hopping is only highlighted for one example hexagon by the dotted arrows. The hopping amplitude along the arrows is $\ii t_2$ and $-\ii t_2$ in the opposite direction ($t_2 \in \mathds{R}$). The polarization direction of the laser field is indicated by the double-headed arrow. The edge of the flake is marked by the red line. Its length, the circumference, is  denoted $s(N)$ in the text. }
        \end{figure}
	
	    Figure \ref{fig1} shows the type of hexagonal Haldanite flakes considered. This kind of flakes has only zigzag edges. Each edge of the flake is formed by $N$ small hexagons, e.g., $N = 4$  in the example in Fig.\ \ref{fig1}.  The circumference of the flake is given by 
	    \begin{equation}\label{eq:s_N}
            s(N) = 6a\left(2N-1\right).
        \end{equation}
        The hopping parameters between nearest and next-nearest neighbors are $t_1$ and $\pm \ii t_2$, respectively. In addition, an alternating on-site potential $M$ ($-M$) is assigned to sublattice site A (B).
	
    	The Hamiltonian reads
    	\begin{align}
    	    \hat{H}_0 &= M \left(\sum_{i\in A} \ket{i}\bra{i} - \sum_{i\in B} \ket{i}\bra{i}\right) + t_1 \sum_{\langle i,j\rangle} \left(\ket{i}\bra{j} + \ket{j}\bra{i}\right) \nonumber\\
    	    &+\sum_{\langle\langle i,j \rangle\rangle}\left(t_2^{ij}\ket{i}\bra{j} + \left(t_2^{ij}\right)^*\ket{j}\bra{i}\right).
    	\end{align}
    	The sum over $i\in A$ ($i\in B$) in the first term includes all sites on sub-lattice  A (B). As usual, $\langle i,j\rangle$ indicates nearest-neighbor pairs and $\langle\langle i,j \rangle\rangle$  next-nearest neighbor pairs. For the next-nearest neighbors the hopping amplitude depends on the direction. The element $t_2^{ij}$ indicates the hopping from site $j$ towards site $i$ (complex conjugate in the opposite direction) and is given by
    	\begin{align}
    	    t_2^{ij} = \begin{cases} \ii t_2 ~~~~~~\mathrm{for~hopping~ along~ the ~arrows}\\-\ii t_2 ~~~~\mathrm{for~hopping~ against~ the~ arrows} 
    	    \end{cases}
    	\end{align}
    	where the arrows are indicated in Fig. \ref{fig1}. 
	    
	   The time-independent Schr\"odinger-equation
	   \begin{equation}
	       \hat{H}_0\ket{\psi_i} = E_i\ket{\psi_i}
	   \end{equation}
	   is solved numerically. The eigenenergies $E_i$ of the respective eigenstates $\ket{\psi_i}$ are sorted in an ascending order, \hbox{$E_0 \leq E_1 \leq E_2 \leq\dots$}.

	   \subsection{Coupling to an external field}

	    The matrix elements of the corresponding  time-dependent Hamiltonian in velocity gauge  are given by \cite{Graf_1995}
	    \begin{align}
	        \bra{i}\hat{H}(t)\ket{j} = \bra{i}\hat{H}_0\ket{j}\ee^{-\ii \left(\bm{r}_{i} - \bm{r}_{j}\right)\cdot \bm{A}(t)},
	    \end{align}
	    where $\bm{r}_{i}$ ($\bm{r}_{j}$) are the coordinates of the sites $i,j$ in position space, and $\bm{A}(t)$ is the vector potential (in dipole approximation).

        States $\ket{\psi_i}$ with an energy smaller than the Fermi energy zero ($E_i <0$) are initially occupied. 
        The time-evolution of a state $\ket{\Psi_i(t)}$ follows the time-dependent Schr\"odinger equation
        \begin{equation}
            \ii\partial_t \ket{\Psi_i(t)} = \hat{H}(t) \ket{\Psi_i(t)},
        \end{equation}
	    which is solved with a fifth-order Dormand-Prince method with adaptive step size control \cite{Numerical_recipies}. The initial states are the eigenstates of the unperturbed system, $\ket{\Psi_i(t = 0)} = \ket{\psi_i}$.
	    
	    \subsection{Calculation of HHG spectra}
	    To obtain the emitted harmonic spectrum, the derivative of the electric current is used.
	    The operator to calculate the current between two sites $i$ and $j$ is given by
	    \begin{align}\label{eq:current_twoSites}
			\hat{\bm{j}}_{i,j}(t) &= 	-\mathrm{i}\left(\bm{r}_{i} - \bm{r}_{j} \right)\nonumber\\
			&\times\left(\ket{i}\bra{i}\hat{H}(t)\ket{j}\bra{j} -  \ket{j}\bra{j}\hat{H}(t)\ket{i}\bra{i}\right).
		\end{align}
	    The contribution of one electron to the total current operator is \cite{Review_Transport}
		\begin{align}\label{eq:current_operator}
			\hat{\bm{j}}(t) &= -\mathrm{i}\sum_{i,j} \hat{\bm{j}}_{i,j}(t)\nonumber \\
			&=-\mathrm{i}\sum_{i,j}\left(\bm{r}_{i} - \bm{r}_{j} \right)\ket{i}\bra{i}\hat{H}(t)\ket{j}\bra{j}.
		\end{align}
	    For the total current, the currents $\bm{j}_l(t)$ due to the electrons in all occupied states $\ket{\Psi_l(t)} $  are summed up,
	    \begin{align}\label{eq:current}
		    \bm{J}(t) = \sum_l \bm{j}_l(t) = \sum_l\bra{\Psi_l(t)}\hat{\bm{j}}(t)\ket{\Psi_l(t)}.
	    \end{align}
	    Fourier-transforming both components of the current, parallel ($\parallel$, $y$-direction) and perpendicular ($\perp$, $x$-direction) to the external field, leads to
	    \begin{align}
	        P_{\parallel,\perp}(\omega) = \left|P_{\parallel,\perp}(\omega)\right|\ee^{\ii\varphi_{\parallel,\perp}(\omega)} =  \int^{+\infty}_{-\infty} \dot{J}_{\parallel,\perp}(t)\, \ee^{-\ii \omega t}\mathrm{d}t.
	    \end{align}
	    The intensity of the emitted light (yield) is proportional to $\left|P_{\parallel,\perp}(\omega)\right|^2$, and its helicity is encoded in the phase difference 
	    \begin{align}\label{eq:phase_difference}
	        \Delta\varphi(\omega) = \varphi_\parallel(\omega) - \varphi_\perp(\omega).
	    \end{align}
	    Numerically, the Fourier-transformation is approximated by a fast Fourier-transformation (FFT). A Hann window is applied before computing the FFT.
	
	   \subsection{Bulk system}
	   
	    The results for HHG in finite flakes are compared to the corresponding  simulation results for Haldanite bulk. How HHG in Haldanite bulk can be calculated is explained in detail in Ref.\ \cite{Moos2020}. The important equations for this work are listed in this section.
	    
	    As usual, a periodic lattice is assumed, which can in this case be broken down to a $2\times2$-Hamiltonian by making a Bloch ansatz. Assuming that the next-nearest neighbor hopping $\ii t_2$ is completely imaginary, one finds
	    \begin{equation}
	        \bm{H}(\bm{k}) = \twobytwo{\sigma}{\tau}{\tau^*}{- \sigma},
        \end{equation}
        with 
        \begin{equation}
            \begin{split}
            	\tau(\bm{k}) &= t_1 \sum_{n=1}^3 \ee^{\ii \bm{\delta}_n \cdot \bm{k}},
            	\\
            	\sigma(\bm{k}) &= M + 2\, t_2 \sum_{n=1}^3 \sin(\bm{g}_n \cdot \bm{k}),
            \end{split}
	    \end{equation}
	    and $\bm{\delta}_n$ and $\bm{g}_n$ being vectors between nearest and next-nearest neighbors, respectively. The time-independent Schr\"odinger equation
	    \begin{align}
	        \bm{H}(\bm{k})\bm{C}_\pm(\bm{k}) = E_\pm(k) \bm{C}_\pm(\bm{k})
	    \end{align}
	    has two solutions for each $\bm{k}$, one for the valence band ($E_-(\bm{k})$, $\bm{C}_-(\bm{k})$) and one for the conduction band ($E_+(\bm{k})$, $\bm{C}_+(\bm{k})$).
	    
	    By coupling the system to an external field $\bm{A}(t)$, the lattice momentum $\bm{k}$ becomes time-dependent
	    \begin{equation}
	        \bm{k}(t) = \bm{k} + \bm{A}(t)
	    \end{equation} 
	    and the time-dependent Schr\"odinger equation becomes
	    \begin{equation}
	        \ii\dot{\bm{C}}(\bm{k},t) = \bm{H}(\bm{k}(t))\bm{C}(\bm{k},t).
	    \end{equation}
	    The initial condition is $\bm{C}(\bm{k}, t = 0) = \bm{C}_-(\bm{k})$.
	    
	    To obtain the harmonic spectra, the current
	    \begin{equation}
	        v_j(\bm{k}, t) = \bm{C}^\dagger(\bm{k},t)  \partial_{k_j} \bm{H} (\bm{k}(t)) \bm{C}(\bm{k},t),
	    \end{equation}
	    is calculated. Here, $j$ stands for the coordinate, it is either $j = x$ or $j = y$.
	    
	    This is calculated for different $\bm{k}$ values within the first Brillouin zone. The total current $v_j$ is obtained by integration over $\bm{k}$. The harmonic spectrum is given by the Fourier-transform of the acceleration $\dot{v}_j$, similar to the finite flakes.  
	    
	    \subsection{Parameters}
		A value of $t_2 = 0.03$ is chosen for the next-nearest neighbor hopping amplitude. The on-site potential is set to $M = 0.01$. Further, the parameters of graphene are used for the distance between nearest neighbors $a=2.68$\,a.u.$\,\simeq\, 1.42$\,\AA\, and for the hopping amplitude between them $t_1 =- 0.1 \, ~$a.u.$\,\simeq \,- 2.7$\,eV \cite{Cooper_2012}. With these values, used throughout the paper,  the system is in the topological phase \cite{Haldane_1988}.
		
		 The linearly polarized laser pulse with a $\sin^2$ envelope is described by the vector potential in dipole approximation
	    \begin{align}
		    \bm{A}(t) = A_0\,\sin ^2\left(\frac{\omega_0 t}{2 n_{0}}\right)\,\sin (\omega_0 t)\, \bm{e}_y
		\end{align}
        for times  $0\leq t \leq 2\pi n_{0}/\omega_0$ and zero otherwise. The number of cycles is set to $n_0 = 10$, the amplitude of the field is $A_0 = 0.05$ (corresponding to an intensity of $\simeq 5\times 10^{9}~ \mathrm{Wcm}^{-2}$), and the angular frequency is  $\omega_0 = 7.5\cdot 10^{-3}$ (wavelength of $\lambda_0 = \SI{6.1}{\micro\meter}$).
	
	\section{Unperturbed hexagonal flakes}\label{sec:static_flakes}

\begin{figure}
        \includegraphics[width = \linewidth]{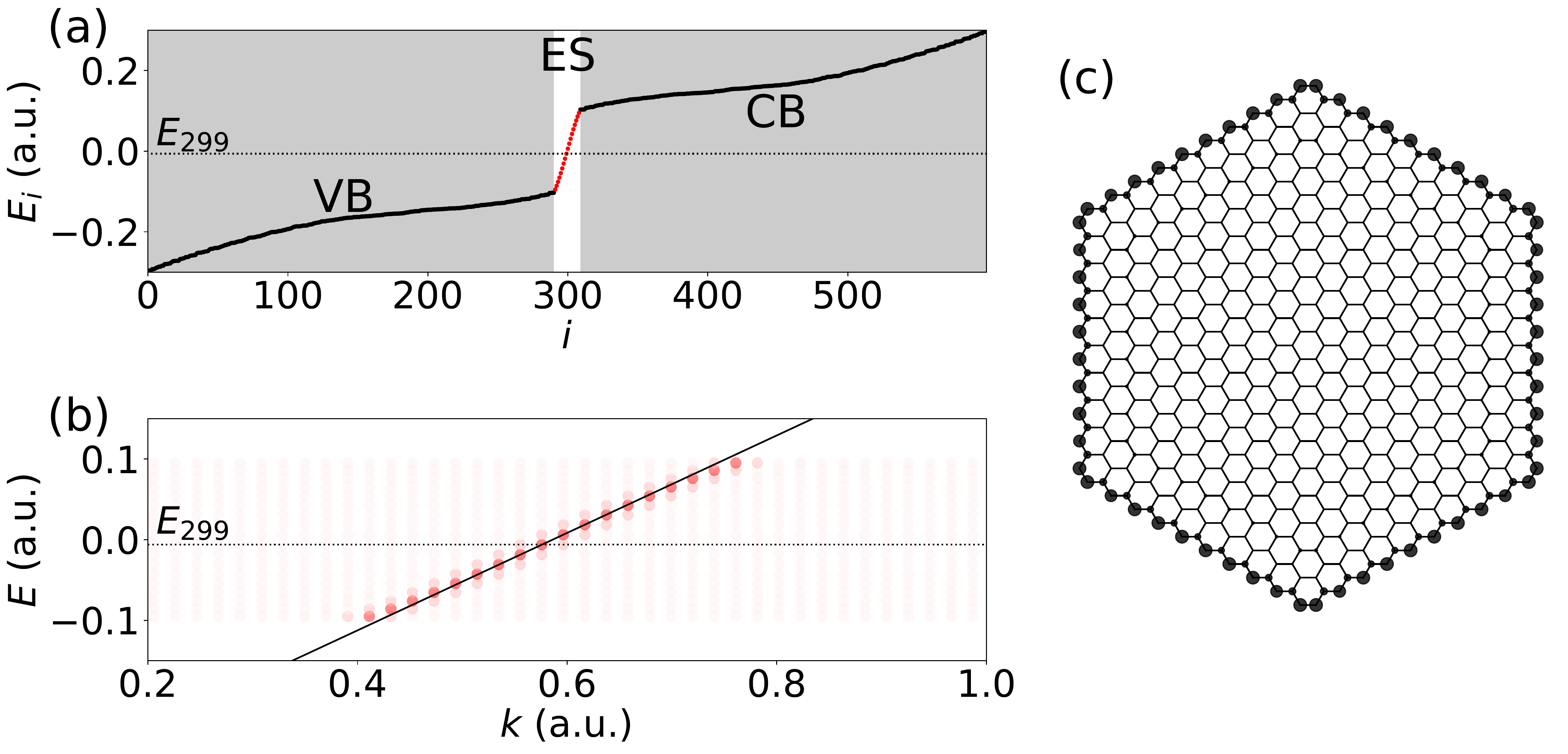}
        \caption{\label{fig2} (a) Energies of a finite flake ($N = 10$) plotted in ascending order. The edge states (ES) are highlighted in red and without a shaded background. The states of the valence band (VB) and conduction band (CB) are indicated by a shaded background. The dotted, horizontal line indicates the energy of the highest occupied state ($E_{299}$). (b) Color-coded Fourier-transform of the edge states. The slope of the solid line is determined by the maxima for the  highest occupied ($i = 299$) and lowest unoccupied ($i = 300$) state.
        (c) Probability density of the highest occupied state ($E_{299}$). The radii of the circles centered at the lattice sites are proportional to the probability density. The localization very close to the edge sites is clearly visible.}
    \end{figure}

	  The energies $E_i$ of the respective eigenstates $\ket{\psi_i}$ of the unperturbed, finite system ($N = 10$) are plotted against the index $i$ in Fig. \ref{fig2}a. The highest occupied state has an energy $E_{299} = -0.00621$.  The electron density of this state is mainly located along the edge of the flake, as shown in Fig. \ref{fig2}c.  There are several such edge states. We call a state an edge state if at least $90\%$ of the probability density is located at the edge of the flake. In Fig. \ref{fig2}a, the edge states (ES) are indicated in red.

    To transform the edge states from position space to momentum space, we Fourier-transform the part of the state that is localized along the edge. The edge is indicated by a red line in Fig. \ref{fig1}, i.e., it is one-dimensional, so the obtained $k$ is also one-dimensional. Each of the transformed states are plotted color-coded in red shades on a linear scale at their energy against momentum $k$. The result for $N = 10$ is depicted in Fig. \ref{fig2}b. The maxima of the absolute value of the Fourier-transformed edge states follow an almost linear dispersion relation in momentum space. The slope of the dispersion relation determines the group velocity $v_g$. To approximate the slope only the two edge states around $E= 0$ are used, i.e., the highest occupied state $\ket{\psi_\mathrm{HO}}$ (for $N=10$ this is $\ket{\psi_{299}}$) and the lowest unoccupied state $\ket{\psi_\mathrm{LU}}$ (for $N=10$ this is $\ket{\psi_{300}}$). The group velocity is
    \begin{equation}\label{eq:v_g}
        v_g = \frac{\partial E}{\partial k} \approx \frac{E_\mathrm{LU} - E_\mathrm{HO}}{k_\mathrm{LU} - k_\mathrm{HO}} =: \frac{\Delta E_\mathrm{edge}}{\Delta k_\mathrm{edge}}.
    \end{equation}
    Here, $k_\mathrm{HO}$ ($k_\mathrm{LU}$) is the $k$-value where the absolute value of the  Fourier-transformed state $\ket{\psi_\mathrm{HO}}$ ($\ket{\psi_\mathrm{LU}}$) has its  maximum.
    
    \begin{figure}
            \includegraphics[width = \linewidth]{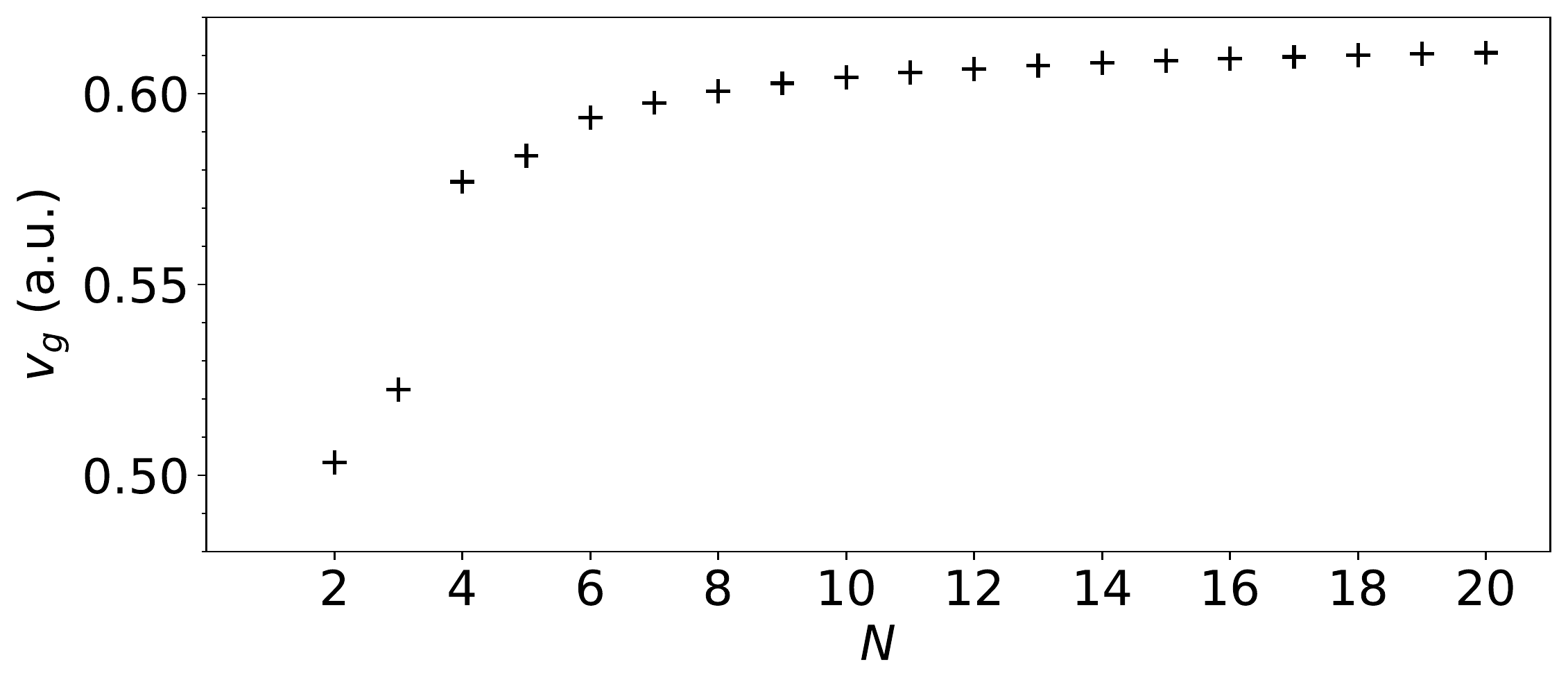}
            \caption{\label{fig3} Group velocity $v_g$ calculated with (\ref{eq:v_g}) as a function of the flake size $N$.}
    \end{figure}
    
    The group velocity $v_g$ defined in this way is calculated for flakes of sizes $N = 1,2,3,...,20$ and  plotted in Fig. \ref{fig3}. For very small flakes the interpretation as a group velocity does not make much sense. Nevertheless, we can formally evaluate \eqref{eq:v_g} and obtain for   $v_g(N = 1) = 0.3312$ (not shown in Fig. \ref{fig3}). The group velocities for $N = 2$ and $N = 3$ are still far off from the values for larger flakes.  The resolution in $k$-space increases with increasing flake size. The further increase of $v_g$ for $N = 4,5,...,20$ is smaller. For the largest considered flake  we find $v_g(N = 20) = 0.6107$.

	\section{Emission spectra}\label{sec:spectra}
	
	    \subsection{Comparison between finite flakes and the bulk system}\label{sec:comp_spectra}

\begin{figure}
            \includegraphics[width = \linewidth]{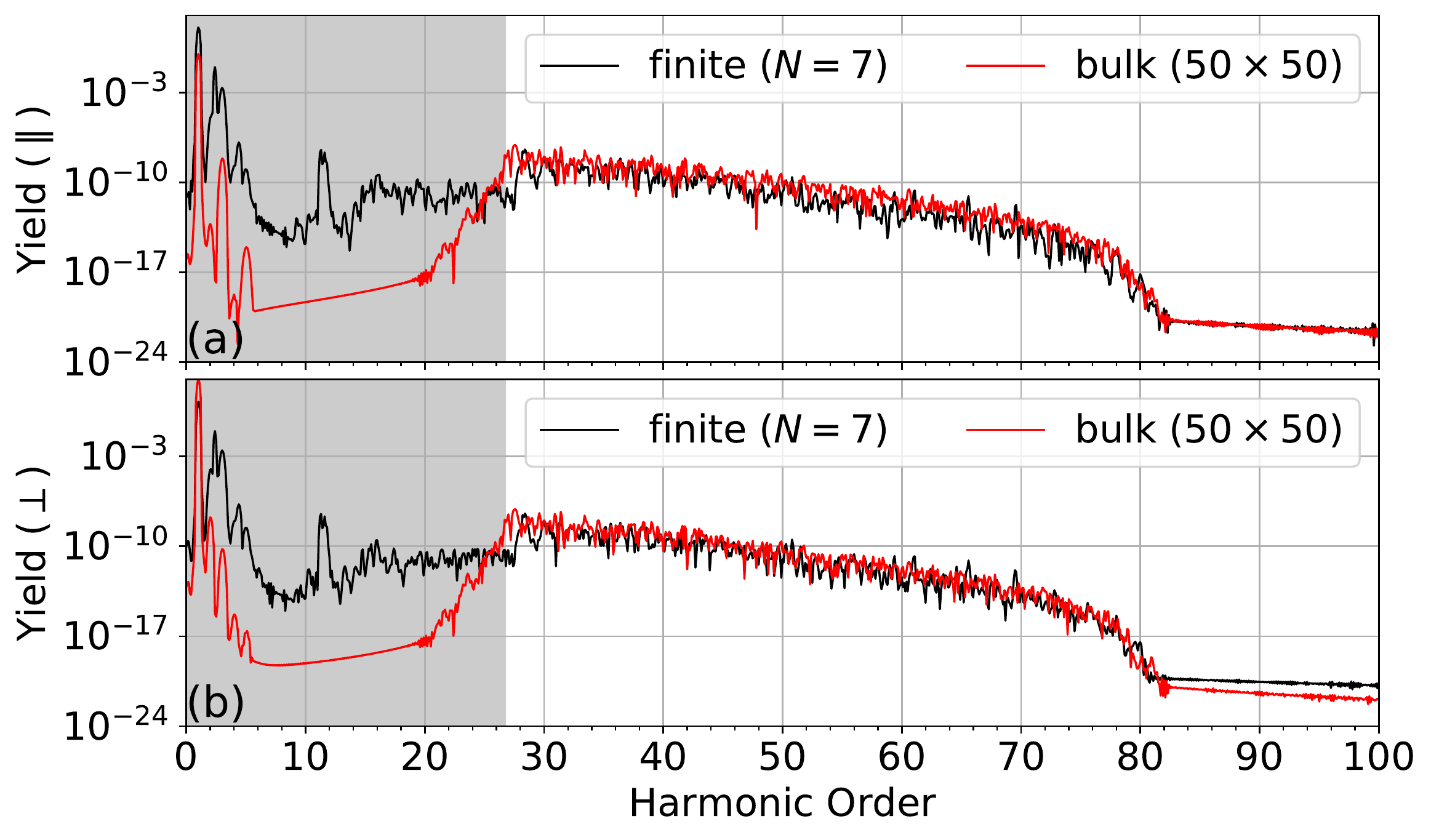}
            \caption{\label{fig4} Comparison between the spectra for the finite ($N=7$) and the bulk ($50\times 50$ points in $k$-space) system in (a) parallel and (b) perpendicular polarization direction to the external laser field. The shaded areas indicate energies below the band gap of the bulk system. The spectrum of the bulk is shifted in yield such that the fundamental harmonic in parallel direction to the external field has a yield equal to $1$. The spectrum of the finite flake is shifted in yield such that the plateaus (non-shaded area) from both systems have a comparable yield in parallel polarization direction.}
        \end{figure}

	   In Fig. \ref{fig4}, the emitted spectra for the finite system ($N = 7$) and the bulk (with a sampling of $50\times50$ points in the first Brillouin zone) are shown. Figure \ref{fig4}a shows the spectra parallel ($y$-direction) and Fig. \ref{fig4}b the spectra perpendicular ($x$-direction) to the polarization direction of the external field. 
	   For energies above the band gap, the spectra for the finite flake and the bulk are similar. 
	   
	   For energies below the band gap, the bulk system shows peaks only at integer harmonic orders up to harmonic $5$. 
	   However, the yield of even harmonics is more enhanced in the perpendicular direction compared to their yield in parallel direction. This originates from the broken symmetry of the flake in $x$-direction (compare sublattice sites on the left and right edge of the flake in Fig. \ref{fig1}). For the finite system, more peaks are observed in that energy region. Due to the presence of edge states in the finite system and the absence of those in the bulk, we expect that at least some of these peaks are related to edge states. Furthermore, the harmonic yield at small energies is for almost every energy smaller in the bulk than in the finite flake (relative to the above band-gap harmonics). This indicates a contribution from the edges of finite flakes to small order harmonics.
	    
	   The intraband harmonics for a fully occupied band usually interfere destructively. This leads to a suppressed harmonic yield for energies below the band gap \cite{bauer_high-harmonic_2018}. This phenomenon can be observed for the bulk as the harmonics below the band gap are suppressed compared to the harmonics above the band gap. For harmonics above the band gap, the interband harmonics start to contribute significantly to the overall spectrum. With edge states present within the band gap, the gap between occupied and unoccupied states effectively decreases. The edge states act as an additional band within the band gap, allowing for emission at photon energies that are not possible for bulk.  It was already shown in previous works that edge states can enhance the harmonic yield for energies below the band gap \cite{bauer_high-harmonic_2018,DrueekeBauer2019,JuerssBauer2019}. A similar effect was found for states within the band gap due to impurities \cite{HHG_doped_solids,HHG_Doped_MgO-Cr}.

	    \subsection{Dependence on the flake size}\label{sec:spectra_size}

\begin{figure}
            \includegraphics[width = \linewidth]{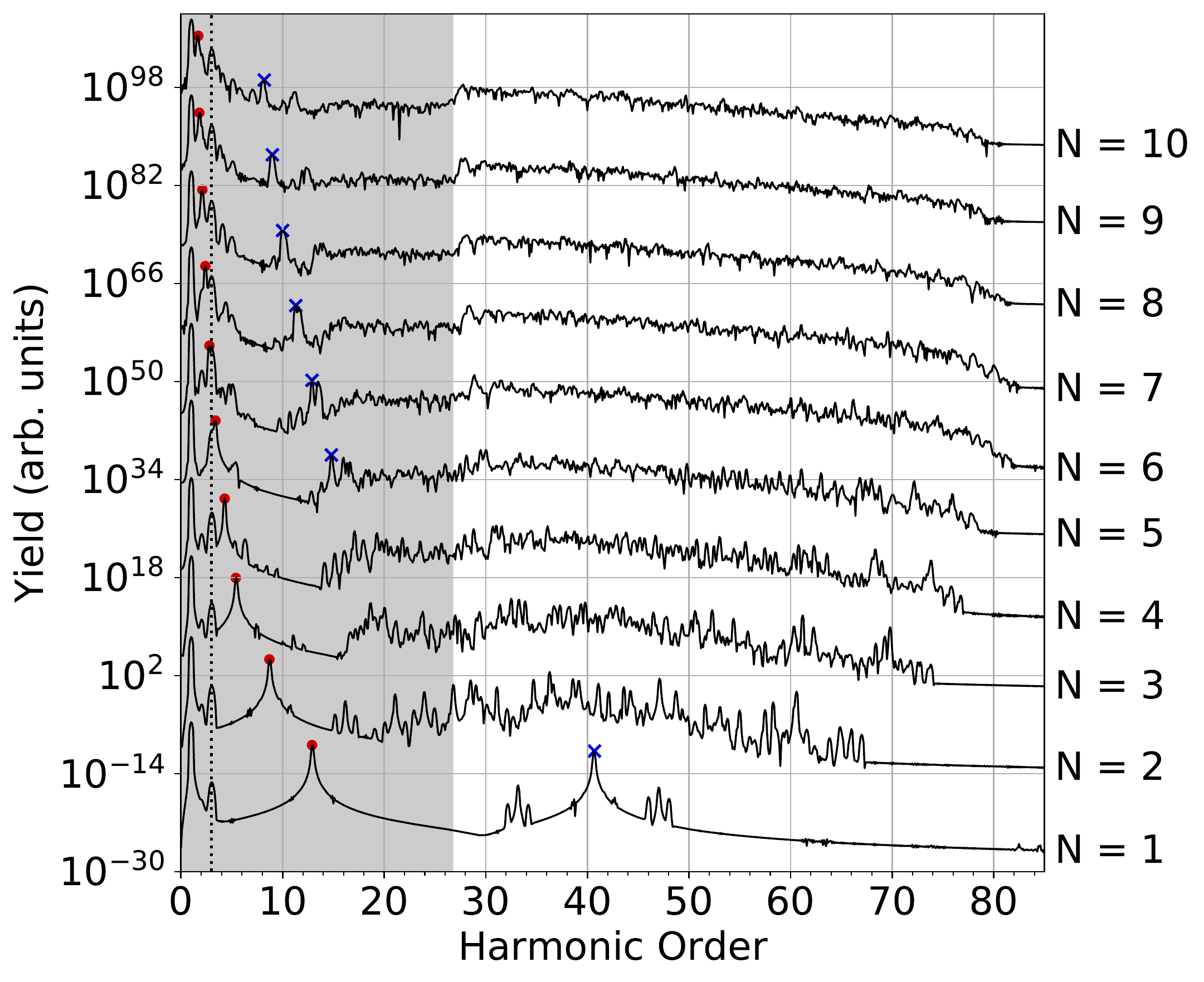}
            \caption{\label{fig5} Harmonic spectra in parallel polarization direction  for different flake sizes $N$. The red bullets (blue crosses) indicate the first (second) peak  that shifts continuously to smaller harmonic orders as the flake size increases. For $N = [2,3,4]$ the second peak could not be unambiguously identified. The spectra are multiplied by a factor of $10^{12 (N-1)}$ for better visibility. The dotted, vertical line indicates harmonic order three.}
        \end{figure}

	    In Fig. \ref{fig5}, the spectra for different flake-sizes ($1\leq N \leq 10$) are plotted. For energies larger than the band gap, the spectra are almost the same for flakes $N \geq  6$. For the harmonics in the sub-band gap regime of the bulk (gray shaded area) the spectra depend on the size of the flakes. In particular, there are two peaks, indicated by a red bullet and a blue cross that can be clearly observed in most of the spectra. These peaks shift towards smaller harmonic orders as the size of the flakes increases. A peak shift related to the carrier envelope phase was observed in Ref. \cite{Schmid2021}. However, the peaks and their shifts that we observe  in this work originate from a different effect, as will be discussed in the following. 
	    
	    Let us denote the position of the first peak (red bullet) by $\omega_1(N)$, the second one (blue cross) by $\omega_2(N)$. Further, the dotted, vertical line indicates harmonic order three in order to see that this common harmonic peak is independent on the flake size $N$. 
	    
	    \subsection{Filtered current}\label{sec:filtered_spectra}
	    
\begin{figure}
            \includegraphics[width = \linewidth]{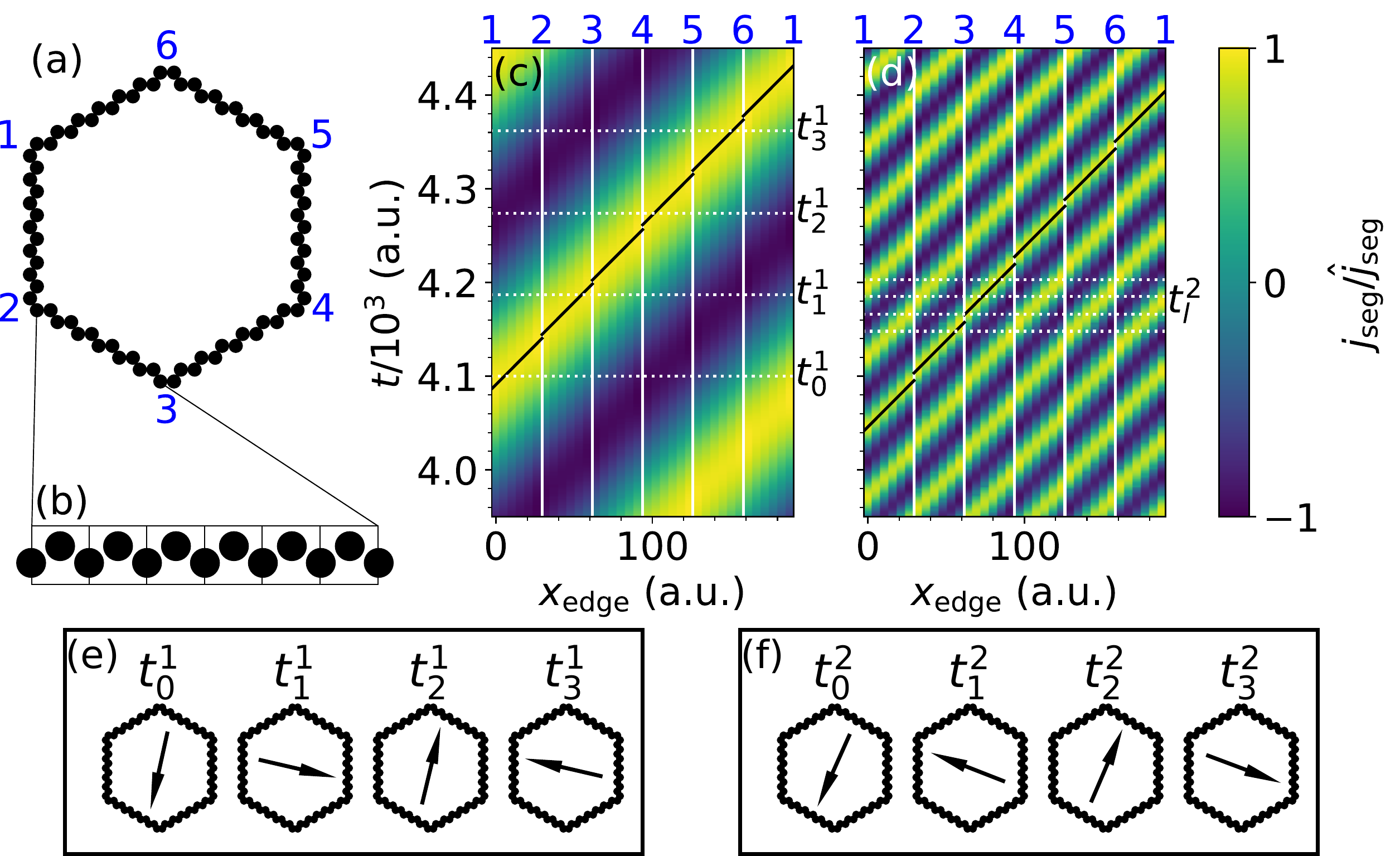}
            \caption{\label{fig6} (a) Hexagonal flake ($N = 7$) with indicated corners (blue numbers). (b) Zoom into one edge (between two corners) with indication of segments containing three sites each. (c and d) Frequency-filtered current, spatially and temporally resolved, for (c)  $\omega_1 = 2.4~\omega_0$, and (d)  $\omega_2 = 11.3~\omega_0$. The white vertical lines in  (c)  and (d) indicate the positions of the corners of the flake. The diagonal black line segments correspond to a velocity $v = 0.5976$, i.e.,  the group velocity determined by the band structure for $N = 7$. These line segments are plotted for each edge between two corners separately. The horizontal lines in (c) and (d) depict different times, $t_l^i$ (with $i = 1$ for the first peak, $i = 2$ for the second one, $l = 0,1,2,3$) which are sorted in an ascending order $t_0^i < t_1^i < t_2^i < t_3^i$. (e and f) Direction of the spatially integrated current (frequency filtered) over the whole edge of the flake depicted as arrow for the first $i = 1$ (e) and second $i = 2$ (f) peak and for different times $t_l^i$. }
        \end{figure}

	    For this consideration we use a flake of size $N = 7$, Fig. \ref{fig6}a. Each edge of the flake (between two corners) is divided into segments consisting of three lattice sites each. This is indicated in Fig. \ref{fig6}b. The total current between the sites within one segment is calculated using  (\ref{eq:current_twoSites}). The corners of the flake are left out because the used segments cannot be placed around those in a consistent way. 
	    
	    To investigate the origin of the peaks, we apply a frequency filter of the form
	    \begin{align}\label{eq:filter}
	        f_m(\omega) = \ee^{-(\omega - \omega_m)^2/(2\sigma^2)}
	    \end{align}
	    to the Fourier-transformed currents of each segment. We chose $\sigma = 0.001$.
	    
	    By dividing the edge into segments, a spatial resolution of the current is achieved. Note, however, that an electron may follow different pathways within one segment. It can either go directly from one edge site of a segment to the opposite edge site  by next-nearest neighbor hopping (hopping distance $\sqrt{3} a$) or sequentially by nearest-neighbor hopping (total hopping distance $2 a$). The latter option has as a larger hopping amplitude so that we use it for the following analysis. It was also used to calculated the  approximate dispersion relation for the edge states shown in Fig. \ref{fig1}. Ignoring the corners, the position of segment $n$ is given by
	    \begin{align}
	        x_\mathrm{edge}(n) = 2an.
	    \end{align}

	    Figures \ref{fig6}c,d show the results for the filtered current, resolved in time and space. The vertical white lines indicate the position of the corners. The numbers of the corners are indicated at the  top of the panels. Their respective positions on the flake are indicated in Fig. \ref{fig6}a. The current is positive if it points in an anti-clockwise direction along the edge of the flake and it is negative if it points in the opposite direction. Because the current around the corners are not captured, the current appears discontinuous at those corner points. Figures \ref{fig6}e,f indicates the direction of the total current of the filtered peak, summed up over the whole edge. The times at which these are plotted, are indicated by the dotted, horizontal lines in Figs. \ref{fig6}c,d.
	    
	    Figure \ref{fig6}c shows the filtered current for $\omega_1 = 2.4~\omega_0$ (first peak) and in Fig. \ref{fig6}d  for  $\omega_2 = 11.3~\omega_0$. The diagonal black lines indicate the positions of particles moving with a velocity of $v = 0.5976$. This velocity is the previously determined group velocity $v_g$ for $N = 7$, calculated from the slope of the edge-state dispersion relation. The black line segments are only plotted between two corners because of the discontinuities at the corners.
	    
	    For the first peak (Fig. \ref{fig6}c) the current has one local maximum and one local minimum along the edge of the chain at all times. The local maximum and minimum are located on the opposite side of the flake for all times. The sign of the current indicates whether the current points in clockwise (negative sign) or anti-clockwise (positive sign) direction along the edge. Hence, the total current is not zero, as shown by the arrow in Fig. \ref{fig6}e. Over time, the total current for this peak rotates in an anti-clockwise direction. The velocity of this rotation is determined by the group velocity $v_g$, which is indicated by the diagonal black lines in Fig. \ref{fig6}c. The change of the total current over times causes the emission of photons.

	    The current filtered around the second peak (Fig. \ref{fig6}d) shows a five-times higher symmetry. There are five local maxima and minima each.  Again, these local maxima and minima move along the edge of the flake with a constant velocity close to $v_g$ (diagonal black lines). The total current for this peak is again finite. Its direction is shown for different times in Fig. \ref{fig6}f. The higher symmetry causes a faster rotation of the total current, and hence photons with a higher energy are emitted. 
	    
	    Note that the rotation of the total current for both peaks is opposite. For the first peak, the total (edge) current rotates in anti-clockwise direction (Fig. \ref{fig6}e) and for the second peak in clockwise direction (Fig. \ref{fig6}f). Comparing the helicities (eq. (\ref{eq:phase_difference})) of the emitted light from both peaks (for all flake sizes) shows that the first peak always emits photons with an opposite helicity compared to the second peak. The values are $-\pi/2$ for the first peak and $\pi/2$ for the second.\newline
	    
	    In the supplemental material \cite{Suppl_material}, a similar calculation for the trivial phase is performed showing that the observed effects are caused by the presence of the topological edge states.

    \section{Connection between sub-band-gap harmonic spectra and edge-state dispersion relation}\label{sec:connection}
	    
	    It was shown in the previous section that the current corresponding to the sub-band-gap peaks rotates with a constant velocity that is very close to the group velocity determined by the edge-state dispersion relation. In this section, we generalize this investigation and include various flake sizes. 
	    
\begin{figure}
            \includegraphics[width = \linewidth]{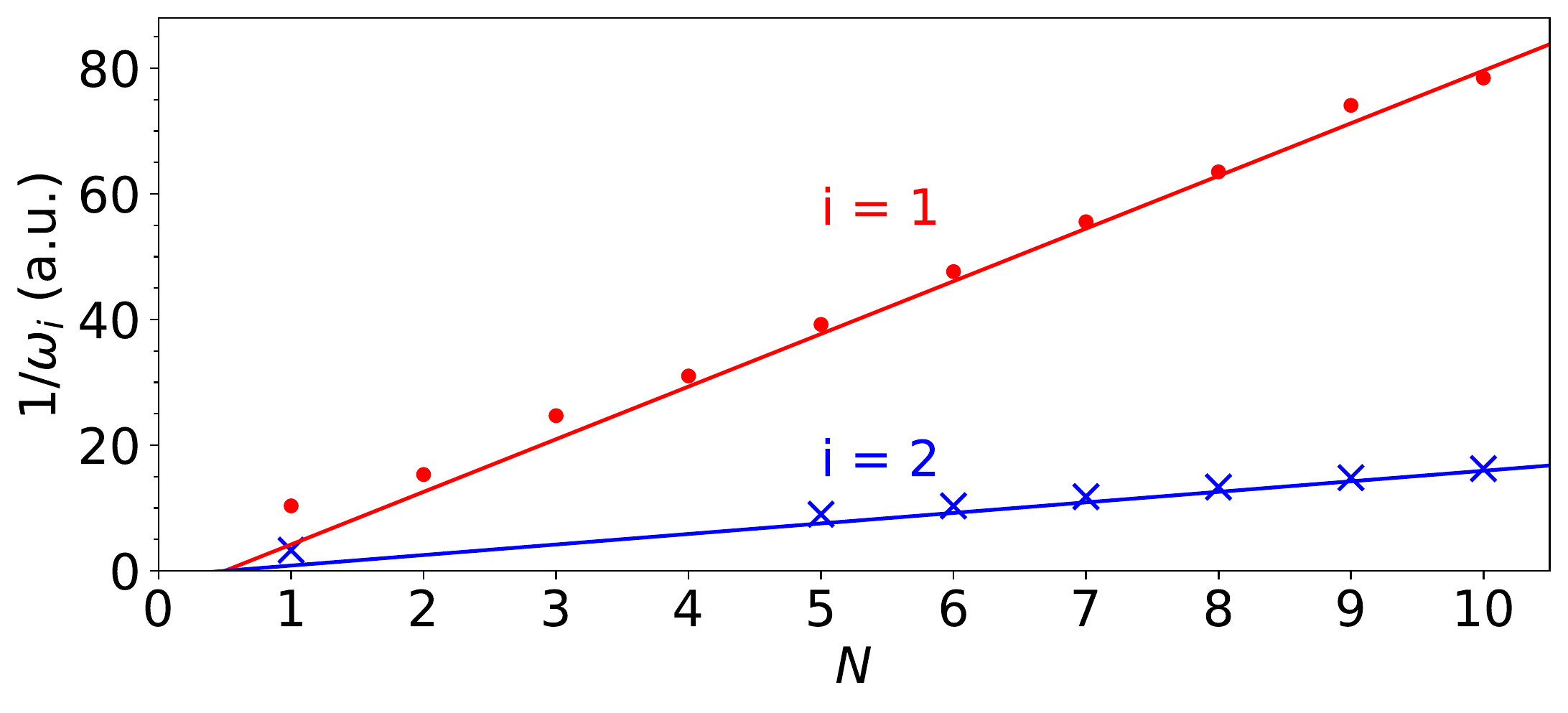}
            \caption{\label{fig7} Data points for the inverse peak positions $1/\omega_i$ for the first ($i = 1$) and second ($i=2$) peak as a function of the flake size $N$.  The red solid line ($i = 1$) corresponds to the theoretical angular frequency $\omega_1^\mathrm{th}$ determined from the group velocity of the edge states. The solid blue line ($i = 2$) corresponds to five-times that group velocity.}
        \end{figure}

	    First, we focus on the $\omega_1$-peak at the smaller harmonic order (indicated with red bullets in Fig. \ref{fig5}). 
	    The circumference $s(N)$ of the flakes depends linearly on $N$, see (\ref{eq:s_N}). In Fig. \ref{fig7}, the red circles indicate the inverse of the peak position $1/\omega_1$ as a function of $N$, showing a linear dependence. This behavior is expected because the group velocity $v_g$ is almost  constant for (sufficiently big)  flake sizes (see Fig. \ref{fig3}),
	    \begin{equation}\label{eq:v_edge}
	        v = \frac{s(N)}{T_1(N)} = \frac{s(N)\, \omega_1^\mathrm{th}(N)}{2\pi} = \mathrm{const.},
	    \end{equation}
	    where $T_1$ is the time the electrons need to encircle the flake and $\omega_1^\mathrm{th}$ is the respective theoretical angular frequency. With
	    \begin{equation}\label{eq:1_omega1}
	        \frac{1}{\omega_1^\mathrm{th}(N)}  = \frac{s(N)}{2\pi\,v}= \frac{6a\,(2N-1)}{2\pi\,v}
	    \end{equation}
	    then follows the linear dependence between $1/{\omega_1^\mathrm{th}(N)}$ and $N$, as seen from the simulations. As the edge-state group velocity $v_g$ (\ref{eq:v_g}) determines the velocity of the movement along the edges we replace $v$ by $v_g$ and obtain  
	    \begin{equation}\label{eq:1_omega1_2}
	        \frac{1}{\omega_1^\mathrm{th}(N)}  = \frac{s(N)}{2\pi\,v_g}= \frac{s(N)\,\Delta k_\mathrm{edge}}{2\pi\,\Delta E_\mathrm{edge}}.
	    \end{equation}
	    Due to the finite numerical resolution  $\Delta k = 2\pi/s$ for the edge states in $k$-space, $\Delta k_\mathrm{edge} = n \Delta k$ where $n = 1, 2, 3, ...$, and thus 
	    \begin{equation}\label{eq:1_omega1_3}
	        \frac{1}{\omega_1^\mathrm{th}(N)}  = \frac{s(N)\,\Delta k_\mathrm{edge}}{2\pi\,\Delta E_\mathrm{edge}} = \frac{n}{\Delta E_\mathrm{edge}}.
	    \end{equation}
	    As expected, we found that $\Delta k _\mathrm{edge} = \Delta k$ (i.e., $n = 1$). This means that the highest occupied and lowest unoccupied edge state are separated in $k$-space by the given resolution, $\Delta k$.
	    Hence, the peak position should be similar to the energy difference between the lowest unoccupied and highest occupied edge state, which was confirmed by our data (not shown). 
	     
	     The red line in Fig. \ref{fig7} was calculated according to (\ref{eq:1_omega1}) with the velocity $v$ set to the group velocity of a large flake  $v_g(N = 20) = 0.6107$, see Fig. \ref{fig3}. The data points agree with the linear function predicted from our theory. The deviation of the data points from the linear function can be explained by the small differences of the numerically determined group velocity for different flake sizes.

	    As seen in the previous section, this indicates that the peaks in the spectra (Fig. \ref{fig5}) marked by the red dots originate from a movement of the current along the edge. The group velocity $v_g$ is similar for sufficiently large flake sizes. Hence, the time for the current to move along the edge of the flake increases with the flake size, and the corresponding frequency decreases. \newline

	    The blue crosses in Fig. \ref{fig7} are the data points for the second peak in the spectra at $\omega_2(N)$. Similar to $\omega_1(N)$, a linear dependence between $1/\omega_2$ and $N$ can be observed but with a smaller slope. It was shown in the previous section that the edge-current corresponding to this peak has a five-times higher symmetry. This leads to a five-times faster oscillation and thus to a peak position $\omega_2$ at a frequency approximately five-times higher than $\omega_1$. The factor between both peaks is not exactly five because of discrete values in our model system (discrete energy spectrum of the unperturbed system, discrete resolution in frequency).
	    The blue line ($i = 2$) is a plot of the linear function
	    \begin{equation}\label{eq:1_omega2}
	        \frac{1}{\omega_2^\mathrm{th}(N)}  = \frac{s(N)}{2\pi\,5\,v_g} =  \frac{6a\,(2N-1)}{2\pi\,5\,v_g}.
	    \end{equation}
	    It agrees very well with the data points. 
	   
	    The observed effect is not restricted to  this kind of hexagonal flakes. We observed similar peaks for triangular flakes. However, the factor between both peaks in the spectra was close to two instead of five in this case. 
	    
    \section{Summary}\label{sec:summary}

        Harmonic spectra for Haldanite bulk and finite, hexagonal Haldanite flakes were compared. Differences occur mainly for harmonic orders below the bulk band gap.  Two strong peaks in the spectra could be identified that depend on the size of the flake. These peaks shift continuously towards smaller harmonic orders as the flake size increases. 
        We could explain the positions of these peaks by non-uniform currents along the edges. The spatial and temporal change in the current causes the emission of light whose frequency is determined by the dispersion relation of the static system. The lower-energy peak corresponds to one round trip of the current's maximum  and also  to the energy difference between lowest unoccupied and highest occupied edge state. The higher-energy peak corresponds to a fifth of a round trip because the envelope of the current has a five times higher symmetry, which leads to a five times faster oscillation in the total, spatially integrated current.

	\section*{Acknowledgment}

	Funding by the German Research Foundation - SFB 1477 “Light-Matter Interactions at Interfaces,” Project No. 441234705, is gratefully acknowledged. C.J. acknowledges financial support by the doctoral fellowship program of the University of Rostock.

	\bibliography{biblio.bib}

\begin{thebibliography}{45}%
\makeatletter
\providecommand \@ifxundefined [1]{%
 \@ifx{#1\undefined}
}%
\providecommand \@ifnum [1]{%
 \ifnum #1\expandafter \@firstoftwo
 \else \expandafter \@secondoftwo
 \fi
}%
\providecommand \@ifx [1]{%
 \ifx #1\expandafter \@firstoftwo
 \else \expandafter \@secondoftwo
 \fi
}%
\providecommand \natexlab [1]{#1}%
\providecommand \enquote  [1]{``#1''}%
\providecommand \bibnamefont  [1]{#1}%
\providecommand \bibfnamefont [1]{#1}%
\providecommand \citenamefont [1]{#1}%
\providecommand \href@noop [0]{\@secondoftwo}%
\providecommand \href [0]{\begingroup \@sanitize@url \@href}%
\providecommand \@href[1]{\@@startlink{#1}\@@href}%
\providecommand \@@href[1]{\endgroup#1\@@endlink}%
\providecommand \@sanitize@url [0]{\catcode `\\12\catcode `\$12\catcode
  `\&12\catcode `\#12\catcode `\^12\catcode `\_12\catcode `\%12\relax}%
\providecommand \@@startlink[1]{}%
\providecommand \@@endlink[0]{}%
\providecommand \url  [0]{\begingroup\@sanitize@url \@url }%
\providecommand \@url [1]{\endgroup\@href {#1}{\urlprefix }}%
\providecommand \urlprefix  [0]{URL }%
\providecommand \Eprint [0]{\href }%
\providecommand \doibase [0]{https://doi.org/}%
\providecommand \selectlanguage [0]{\@gobble}%
\providecommand \bibinfo  [0]{\@secondoftwo}%
\providecommand \bibfield  [0]{\@secondoftwo}%
\providecommand \translation [1]{[#1]}%
\providecommand \BibitemOpen [0]{}%
\providecommand \bibitemStop [0]{}%
\providecommand \bibitemNoStop [0]{.\EOS\space}%
\providecommand \EOS [0]{\spacefactor3000\relax}%
\providecommand \BibitemShut  [1]{\csname bibitem#1\endcsname}%
\let\auto@bib@innerbib\@empty
\bibitem [{\citenamefont {Ghimire}\ \emph {et~al.}(2011)\citenamefont
  {Ghimire}, \citenamefont {DiChiara}, \citenamefont {Sistrunk}, \citenamefont
  {Agostini}, \citenamefont {DiMauro},\ and\ \citenamefont
  {Reis}}]{Ghimire2011}%
  \BibitemOpen
  \bibfield  {author} {\bibinfo {author} {\bibfnamefont {S.}~\bibnamefont
  {Ghimire}}, \bibinfo {author} {\bibfnamefont {A.~D.}\ \bibnamefont
  {DiChiara}}, \bibinfo {author} {\bibfnamefont {E.}~\bibnamefont {Sistrunk}},
  \bibinfo {author} {\bibfnamefont {P.}~\bibnamefont {Agostini}}, \bibinfo
  {author} {\bibfnamefont {L.~F.}\ \bibnamefont {DiMauro}},\ and\ \bibinfo
  {author} {\bibfnamefont {D.~A.}\ \bibnamefont {Reis}},\ }\bibfield  {title}
  {\bibinfo {title} {Observation of high-order harmonic generation in a bulk
  crystal},\ }\href {https://doi.org/10.1038/nphys1847} {\bibfield  {journal}
  {\bibinfo  {journal} {Nat Phys}\ }\textbf {\bibinfo {volume} {7}},\ \bibinfo
  {pages} {138} (\bibinfo {year} {2011})}\BibitemShut {NoStop}%
\bibitem [{\citenamefont {Vampa}\ \emph {et~al.}(2015)\citenamefont {Vampa},
  \citenamefont {Hammond}, \citenamefont {Thir\'e}, \citenamefont {Schmidt},
  \citenamefont {L\'egar\'e}, \citenamefont {McDonald}, \citenamefont {Brabec},
  \citenamefont {Klug},\ and\ \citenamefont
  {Corkum}}]{VampaPhysRevLett.115.193603}%
  \BibitemOpen
  \bibfield  {author} {\bibinfo {author} {\bibfnamefont {G.}~\bibnamefont
  {Vampa}}, \bibinfo {author} {\bibfnamefont {T.~J.}\ \bibnamefont {Hammond}},
  \bibinfo {author} {\bibfnamefont {N.}~\bibnamefont {Thir\'e}}, \bibinfo
  {author} {\bibfnamefont {B.~E.}\ \bibnamefont {Schmidt}}, \bibinfo {author}
  {\bibfnamefont {F.}~\bibnamefont {L\'egar\'e}}, \bibinfo {author}
  {\bibfnamefont {C.~R.}\ \bibnamefont {McDonald}}, \bibinfo {author}
  {\bibfnamefont {T.}~\bibnamefont {Brabec}}, \bibinfo {author} {\bibfnamefont
  {D.~D.}\ \bibnamefont {Klug}},\ and\ \bibinfo {author} {\bibfnamefont
  {P.~B.}\ \bibnamefont {Corkum}},\ }\bibfield  {title} {\bibinfo {title}
  {All-optical reconstruction of crystal band structure},\ }\href
  {https://doi.org/10.1103/PhysRevLett.115.193603} {\bibfield  {journal}
  {\bibinfo  {journal} {Phys. Rev. Lett.}\ }\textbf {\bibinfo {volume} {115}},\
  \bibinfo {pages} {193603} (\bibinfo {year} {2015})}\BibitemShut {NoStop}%
\bibitem [{\citenamefont {Tancogne-Dejean}\ \emph {et~al.}(2017)\citenamefont
  {Tancogne-Dejean}, \citenamefont {M\"ucke}, \citenamefont {K\"artner},\ and\
  \citenamefont {Rubio}}]{TancPhysRevLett.118.087403}%
  \BibitemOpen
  \bibfield  {author} {\bibinfo {author} {\bibfnamefont {N.}~\bibnamefont
  {Tancogne-Dejean}}, \bibinfo {author} {\bibfnamefont {O.~D.}\ \bibnamefont
  {M\"ucke}}, \bibinfo {author} {\bibfnamefont {F.~X.}\ \bibnamefont
  {K\"artner}},\ and\ \bibinfo {author} {\bibfnamefont {A.}~\bibnamefont
  {Rubio}},\ }\bibfield  {title} {\bibinfo {title} {Impact of the electronic
  band structure in high-harmonic generation spectra of solids},\ }\href
  {https://doi.org/10.1103/PhysRevLett.118.087403} {\bibfield  {journal}
  {\bibinfo  {journal} {Phys. Rev. Lett.}\ }\textbf {\bibinfo {volume} {118}},\
  \bibinfo {pages} {087403} (\bibinfo {year} {2017})}\BibitemShut {NoStop}%
\bibitem [{\citenamefont {Lakhotia}\ \emph {et~al.}(2020)\citenamefont
  {Lakhotia}, \citenamefont {Kim}, \citenamefont {Zhan}, \citenamefont {Hu},
  \citenamefont {Meng},\ and\ \citenamefont {Goulielmakis}}]{Lakhotia2020}%
  \BibitemOpen
  \bibfield  {author} {\bibinfo {author} {\bibfnamefont {H.}~\bibnamefont
  {Lakhotia}}, \bibinfo {author} {\bibfnamefont {H.~Y.}\ \bibnamefont {Kim}},
  \bibinfo {author} {\bibfnamefont {M.}~\bibnamefont {Zhan}}, \bibinfo {author}
  {\bibfnamefont {S.}~\bibnamefont {Hu}}, \bibinfo {author} {\bibfnamefont
  {S.}~\bibnamefont {Meng}},\ and\ \bibinfo {author} {\bibfnamefont
  {E.}~\bibnamefont {Goulielmakis}},\ }\bibfield  {title} {\bibinfo {title}
  {Laser picoscopy of valence electrons in solids},\ }\href
  {https://doi.org/10.1038/s41586-020-2429-z} {\bibfield  {journal} {\bibinfo
  {journal} {Nature}\ }\textbf {\bibinfo {volume} {583}},\ \bibinfo {pages}
  {55} (\bibinfo {year} {2020})}\BibitemShut {NoStop}%
\bibitem [{\citenamefont {Schubert}\ \emph {et~al.}(2014)\citenamefont
  {Schubert}, \citenamefont {Hohenleutner}, \citenamefont {Langer},
  \citenamefont {Urbanek}, \citenamefont {Lange}, \citenamefont {Huttner},
  \citenamefont {Golde}, \citenamefont {Meier}, \citenamefont {Kira},
  \citenamefont {Koch},\ and\ \citenamefont {Huber}}]{SchubertO.2014}%
  \BibitemOpen
  \bibfield  {author} {\bibinfo {author} {\bibfnamefont {O.}~\bibnamefont
  {Schubert}}, \bibinfo {author} {\bibfnamefont {M.}~\bibnamefont
  {Hohenleutner}}, \bibinfo {author} {\bibfnamefont {F.}~\bibnamefont
  {Langer}}, \bibinfo {author} {\bibfnamefont {B.}~\bibnamefont {Urbanek}},
  \bibinfo {author} {\bibfnamefont {C.}~\bibnamefont {Lange}}, \bibinfo
  {author} {\bibfnamefont {U.}~\bibnamefont {Huttner}}, \bibinfo {author}
  {\bibfnamefont {D.}~\bibnamefont {Golde}}, \bibinfo {author} {\bibfnamefont
  {T.}~\bibnamefont {Meier}}, \bibinfo {author} {\bibfnamefont
  {M.}~\bibnamefont {Kira}}, \bibinfo {author} {\bibfnamefont {S.}~\bibnamefont
  {Koch}},\ and\ \bibinfo {author} {\bibfnamefont {R.}~\bibnamefont {Huber}},\
  }\bibfield  {title} {\bibinfo {title} {Sub-cycle control of terahertz
  high-harmonic generation by dynamical {Bloch} oscillations},\ }\href
  {http://dx.doi.org/10.1038/nphoton.2013.349} {\bibfield  {journal} {\bibinfo
  {journal} {Nat Photon}\ }\textbf {\bibinfo {volume} {8}},\ \bibinfo {pages}
  {119} (\bibinfo {year} {2014})}\BibitemShut {NoStop}%
\bibitem [{\citenamefont {Hohenleutner}\ \emph {et~al.}(2015)\citenamefont
  {Hohenleutner}, \citenamefont {Langer}, \citenamefont {Schubert},
  \citenamefont {Knorr}, \citenamefont {Huttner}, \citenamefont {Koch},
  \citenamefont {Kira},\ and\ \citenamefont {Huber}}]{Hohenleutner2015}%
  \BibitemOpen
  \bibfield  {author} {\bibinfo {author} {\bibfnamefont {M.}~\bibnamefont
  {Hohenleutner}}, \bibinfo {author} {\bibfnamefont {F.}~\bibnamefont
  {Langer}}, \bibinfo {author} {\bibfnamefont {O.}~\bibnamefont {Schubert}},
  \bibinfo {author} {\bibfnamefont {M.}~\bibnamefont {Knorr}}, \bibinfo
  {author} {\bibfnamefont {U.}~\bibnamefont {Huttner}}, \bibinfo {author}
  {\bibfnamefont {S.~W.}\ \bibnamefont {Koch}}, \bibinfo {author}
  {\bibfnamefont {M.}~\bibnamefont {Kira}},\ and\ \bibinfo {author}
  {\bibfnamefont {R.}~\bibnamefont {Huber}},\ }\bibfield  {title} {\bibinfo
  {title} {Real-time observation of interfering crystal electrons in
  high-harmonic generation},\ }\href {http://dx.doi.org/10.1038/nature14652}
  {\bibfield  {journal} {\bibinfo  {journal} {Nature}\ }\textbf {\bibinfo
  {volume} {523}},\ \bibinfo {pages} {572} (\bibinfo {year}
  {2015})}\BibitemShut {NoStop}%
\bibitem [{\citenamefont {You}\ \emph {et~al.}(2017)\citenamefont {You},
  \citenamefont {Yin}, \citenamefont {Wu}, \citenamefont {Chew}, \citenamefont
  {Ren}, \citenamefont {Zhuang}, \citenamefont {Gholam-Mirzaei}, \citenamefont
  {Chini}, \citenamefont {Chang},\ and\ \citenamefont {Ghimire}}]{You2017}%
  \BibitemOpen
  \bibfield  {author} {\bibinfo {author} {\bibfnamefont {Y.~S.}\ \bibnamefont
  {You}}, \bibinfo {author} {\bibfnamefont {Y.}~\bibnamefont {Yin}}, \bibinfo
  {author} {\bibfnamefont {Y.}~\bibnamefont {Wu}}, \bibinfo {author}
  {\bibfnamefont {A.}~\bibnamefont {Chew}}, \bibinfo {author} {\bibfnamefont
  {X.}~\bibnamefont {Ren}}, \bibinfo {author} {\bibfnamefont {F.}~\bibnamefont
  {Zhuang}}, \bibinfo {author} {\bibfnamefont {S.}~\bibnamefont
  {Gholam-Mirzaei}}, \bibinfo {author} {\bibfnamefont {M.}~\bibnamefont
  {Chini}}, \bibinfo {author} {\bibfnamefont {Z.}~\bibnamefont {Chang}},\ and\
  \bibinfo {author} {\bibfnamefont {S.}~\bibnamefont {Ghimire}},\ }\bibfield
  {title} {\bibinfo {title} {High-harmonic generation in amorphous solids},\
  }\href {https://doi.org/10.1038/s41467-017-00989-4} {\bibfield  {journal}
  {\bibinfo  {journal} {Nature Communications}\ }\textbf {\bibinfo {volume}
  {8}},\ \bibinfo {pages} {724} (\bibinfo {year} {2017})}\BibitemShut {NoStop}%
\bibitem [{\citenamefont {Baudisch}\ \emph {et~al.}(2018)\citenamefont
  {Baudisch}, \citenamefont {Marini}, \citenamefont {Cox}, \citenamefont {Zhu},
  \citenamefont {Silva}, \citenamefont {Teichmann}, \citenamefont {Massicotte},
  \citenamefont {Koppens}, \citenamefont {Levitov}, \citenamefont
  {Garc{\'i}a~de Abajo},\ and\ \citenamefont {Biegert}}]{Baudisch2018}%
  \BibitemOpen
  \bibfield  {author} {\bibinfo {author} {\bibfnamefont {M.}~\bibnamefont
  {Baudisch}}, \bibinfo {author} {\bibfnamefont {A.}~\bibnamefont {Marini}},
  \bibinfo {author} {\bibfnamefont {J.~D.}\ \bibnamefont {Cox}}, \bibinfo
  {author} {\bibfnamefont {T.}~\bibnamefont {Zhu}}, \bibinfo {author}
  {\bibfnamefont {F.}~\bibnamefont {Silva}}, \bibinfo {author} {\bibfnamefont
  {S.}~\bibnamefont {Teichmann}}, \bibinfo {author} {\bibfnamefont
  {M.}~\bibnamefont {Massicotte}}, \bibinfo {author} {\bibfnamefont
  {F.}~\bibnamefont {Koppens}}, \bibinfo {author} {\bibfnamefont {L.~S.}\
  \bibnamefont {Levitov}}, \bibinfo {author} {\bibfnamefont {F.~J.}\
  \bibnamefont {Garc{\'i}a~de Abajo}},\ and\ \bibinfo {author} {\bibfnamefont
  {J.}~\bibnamefont {Biegert}},\ }\bibfield  {title} {\bibinfo {title}
  {{Ultrafast nonlinear optical response of Dirac fermions in graphene}},\
  }\href {https://doi.org/10.1038/s41467-018-03413-7} {\bibfield  {journal}
  {\bibinfo  {journal} {Nature Communications}\ }\textbf {\bibinfo {volume}
  {9}},\ \bibinfo {pages} {1018} (\bibinfo {year} {2018})}\BibitemShut
  {NoStop}%
\bibitem [{\citenamefont {Luu}\ \emph {et~al.}(2015)\citenamefont {Luu},
  \citenamefont {Garg}, \citenamefont {Kruchinin}, \citenamefont {Moulet},
  \citenamefont {Hassan},\ and\ \citenamefont {Goulielmakis}}]{Luu2015}%
  \BibitemOpen
  \bibfield  {author} {\bibinfo {author} {\bibfnamefont {T.~T.}\ \bibnamefont
  {Luu}}, \bibinfo {author} {\bibfnamefont {M.}~\bibnamefont {Garg}}, \bibinfo
  {author} {\bibfnamefont {S.~Y.}\ \bibnamefont {Kruchinin}}, \bibinfo {author}
  {\bibfnamefont {A.}~\bibnamefont {Moulet}}, \bibinfo {author} {\bibfnamefont
  {M.~T.}\ \bibnamefont {Hassan}},\ and\ \bibinfo {author} {\bibfnamefont
  {E.}~\bibnamefont {Goulielmakis}},\ }\bibfield  {title} {\bibinfo {title}
  {Extreme ultraviolet high-harmonic spectroscopy of solids},\ }\href
  {http://dx.doi.org/10.1038/nature14456} {\bibfield  {journal} {\bibinfo
  {journal} {Nature}\ }\textbf {\bibinfo {volume} {521}},\ \bibinfo {pages}
  {498} (\bibinfo {year} {2015})}\BibitemShut {NoStop}%
\bibitem [{\citenamefont {Ndabashimiye}\ \emph {et~al.}(2016)\citenamefont
  {Ndabashimiye}, \citenamefont {Ghimire}, \citenamefont {Wu}, \citenamefont
  {Browne}, \citenamefont {Schafer}, \citenamefont {Gaarde},\ and\
  \citenamefont {Reis}}]{Ndabashimiye2016}%
  \BibitemOpen
  \bibfield  {author} {\bibinfo {author} {\bibfnamefont {G.}~\bibnamefont
  {Ndabashimiye}}, \bibinfo {author} {\bibfnamefont {S.}~\bibnamefont
  {Ghimire}}, \bibinfo {author} {\bibfnamefont {M.}~\bibnamefont {Wu}},
  \bibinfo {author} {\bibfnamefont {D.~A.}\ \bibnamefont {Browne}}, \bibinfo
  {author} {\bibfnamefont {K.~J.}\ \bibnamefont {Schafer}}, \bibinfo {author}
  {\bibfnamefont {M.~B.}\ \bibnamefont {Gaarde}},\ and\ \bibinfo {author}
  {\bibfnamefont {D.~A.}\ \bibnamefont {Reis}},\ }\bibfield  {title} {\bibinfo
  {title} {Solid-state harmonics beyond the atomic limit},\ }\href
  {http://dx.doi.org/10.1038/nature17660} {\bibfield  {journal} {\bibinfo
  {journal} {Nature}\ }\textbf {\bibinfo {volume} {534}},\ \bibinfo {pages}
  {520} (\bibinfo {year} {2016})}\BibitemShut {NoStop}%
\bibitem [{\citenamefont {Langer}\ \emph {et~al.}(2017)\citenamefont {Langer},
  \citenamefont {Hohenleutner}, \citenamefont {Huttner}, \citenamefont {Koch},
  \citenamefont {Kira},\ and\ \citenamefont {Huber}}]{LangerF.2017}%
  \BibitemOpen
  \bibfield  {author} {\bibinfo {author} {\bibfnamefont {F.}~\bibnamefont
  {Langer}}, \bibinfo {author} {\bibfnamefont {M.}~\bibnamefont
  {Hohenleutner}}, \bibinfo {author} {\bibfnamefont {U.}~\bibnamefont
  {Huttner}}, \bibinfo {author} {\bibfnamefont {S.}~\bibnamefont {Koch}},
  \bibinfo {author} {\bibfnamefont {M.}~\bibnamefont {Kira}},\ and\ \bibinfo
  {author} {\bibfnamefont {R.}~\bibnamefont {Huber}},\ }\bibfield  {title}
  {\bibinfo {title} {Symmetry-controlled temporal structure of high-harmonic
  carrier fields from a bulk crystal},\ }\href
  {http://dx.doi.org/10.1038/nphoton.2017.29} {\bibfield  {journal} {\bibinfo
  {journal} {Nat Photon}\ }\textbf {\bibinfo {volume} {11}},\ \bibinfo {pages}
  {227} (\bibinfo {year} {2017})}\BibitemShut {NoStop}%
\bibitem [{\citenamefont {Vampa}\ and\ \citenamefont
  {Brabec}(2017)}]{vampa_merge_2017}%
  \BibitemOpen
  \bibfield  {author} {\bibinfo {author} {\bibfnamefont {G.}~\bibnamefont
  {Vampa}}\ and\ \bibinfo {author} {\bibfnamefont {T.}~\bibnamefont {Brabec}},\
  }\bibfield  {title} {\bibinfo {title} {Merge of high harmonic generation from
  gases and solids and its implications for attosecond science},\ }\href
  {https://doi.org/10.1088/1361-6455/aa528d} {\bibfield  {journal} {\bibinfo
  {journal} {Journal of Physics B: Atomic, Molecular and Optical Physics}\
  }\textbf {\bibinfo {volume} {50}},\ \bibinfo {pages} {083001} (\bibinfo
  {year} {2017})}\BibitemShut {NoStop}%
\bibitem [{\citenamefont {Corkum}(1993)}]{corkum_plasma_1993}%
  \BibitemOpen
  \bibfield  {author} {\bibinfo {author} {\bibfnamefont {P.~B.}\ \bibnamefont
  {Corkum}},\ }\bibfield  {title} {\bibinfo {title} {Plasma perspective on
  strong field multiphoton ionization},\ }\href
  {https://doi.org/10.1103/PhysRevLett.71.1994} {\bibfield  {journal} {\bibinfo
   {journal} {Physical Review Letters}\ }\textbf {\bibinfo {volume} {71}},\
  \bibinfo {pages} {1994} (\bibinfo {year} {1993})}\BibitemShut {NoStop}%
\bibitem [{\citenamefont {Lewenstein}\ \emph {et~al.}(1994)\citenamefont
  {Lewenstein}, \citenamefont {Balcou}, \citenamefont {Ivanov}, \citenamefont
  {L'Huillier},\ and\ \citenamefont {Corkum}}]{LewensteinPhysRevA.49.2117}%
  \BibitemOpen
  \bibfield  {author} {\bibinfo {author} {\bibfnamefont {M.}~\bibnamefont
  {Lewenstein}}, \bibinfo {author} {\bibfnamefont {P.}~\bibnamefont {Balcou}},
  \bibinfo {author} {\bibfnamefont {M.~Y.}\ \bibnamefont {Ivanov}}, \bibinfo
  {author} {\bibfnamefont {A.}~\bibnamefont {L'Huillier}},\ and\ \bibinfo
  {author} {\bibfnamefont {P.~B.}\ \bibnamefont {Corkum}},\ }\bibfield  {title}
  {\bibinfo {title} {Theory of high-harmonic generation by low-frequency laser
  fields},\ }\href {https://doi.org/10.1103/PhysRevA.49.2117} {\bibfield
  {journal} {\bibinfo  {journal} {Phys. Rev. A}\ }\textbf {\bibinfo {volume}
  {49}},\ \bibinfo {pages} {2117} (\bibinfo {year} {1994})}\BibitemShut
  {NoStop}%
\bibitem [{\citenamefont {Yue}\ and\ \citenamefont
  {Gaarde}(2021)}]{Gaarde_imperfect_recol}%
  \BibitemOpen
  \bibfield  {author} {\bibinfo {author} {\bibfnamefont {L.}~\bibnamefont
  {Yue}}\ and\ \bibinfo {author} {\bibfnamefont {M.~B.}\ \bibnamefont
  {Gaarde}},\ }\bibfield  {title} {\bibinfo {title} {{Expanded view of
  electron-hole recollisions in solid-state high-order harmonic generation:
  Full-Brillouin-zone tunneling and imperfect recollisions}},\ }\href
  {https://doi.org/10.1103/PhysRevA.103.063105} {\bibfield  {journal} {\bibinfo
   {journal} {Phys. Rev. A}\ }\textbf {\bibinfo {volume} {103}},\ \bibinfo
  {pages} {063105} (\bibinfo {year} {2021})}\BibitemShut {NoStop}%
\bibitem [{\citenamefont {J{\"u}rgens}\ \emph {et~al.}(2020)\citenamefont
  {J{\"u}rgens}, \citenamefont {Liewehr}, \citenamefont {Kruse}, \citenamefont
  {Peltz}, \citenamefont {Engel}, \citenamefont {Husakou}, \citenamefont
  {Witting}, \citenamefont {Ivanov}, \citenamefont {Vrakking}, \citenamefont
  {Fennel},\ and\ \citenamefont {Mermillod-Blondin}}]{Jurgens2020}%
  \BibitemOpen
  \bibfield  {author} {\bibinfo {author} {\bibfnamefont {P.}~\bibnamefont
  {J{\"u}rgens}}, \bibinfo {author} {\bibfnamefont {B.}~\bibnamefont
  {Liewehr}}, \bibinfo {author} {\bibfnamefont {B.}~\bibnamefont {Kruse}},
  \bibinfo {author} {\bibfnamefont {C.}~\bibnamefont {Peltz}}, \bibinfo
  {author} {\bibfnamefont {D.}~\bibnamefont {Engel}}, \bibinfo {author}
  {\bibfnamefont {A.}~\bibnamefont {Husakou}}, \bibinfo {author} {\bibfnamefont
  {T.}~\bibnamefont {Witting}}, \bibinfo {author} {\bibfnamefont
  {M.}~\bibnamefont {Ivanov}}, \bibinfo {author} {\bibfnamefont {M.~J.~J.}\
  \bibnamefont {Vrakking}}, \bibinfo {author} {\bibfnamefont {T.}~\bibnamefont
  {Fennel}},\ and\ \bibinfo {author} {\bibfnamefont {A.}~\bibnamefont
  {Mermillod-Blondin}},\ }\bibfield  {title} {\bibinfo {title} {Origin of
  strong-field-induced low-order harmonic generation in amorphous quartz},\
  }\href {https://doi.org/10.1038/s41567-020-0943-4} {\bibfield  {journal}
  {\bibinfo  {journal} {Nature Physics}\ }\textbf {\bibinfo {volume} {16}},\
  \bibinfo {pages} {1035} (\bibinfo {year} {2020})}\BibitemShut {NoStop}%
\bibitem [{\citenamefont {Yue}\ and\ \citenamefont
  {Gaarde}(2022)}]{yue2021introduction}%
  \BibitemOpen
  \bibfield  {author} {\bibinfo {author} {\bibfnamefont {L.}~\bibnamefont
  {Yue}}\ and\ \bibinfo {author} {\bibfnamefont {M.~B.}\ \bibnamefont
  {Gaarde}},\ }\bibfield  {title} {\bibinfo {title} {Introduction to theory of
  high-harmonic generation in solids: tutorial},\ }\href
  {https://doi.org/10.1364/JOSAB.448602} {\bibfield  {journal} {\bibinfo
  {journal} {J. Opt. Soc. Am. B}\ }\textbf {\bibinfo {volume} {39}},\ \bibinfo
  {pages} {535} (\bibinfo {year} {2022})}\BibitemShut {NoStop}%
\bibitem [{\citenamefont {Goulielmakis}\ and\ \citenamefont
  {Brabec}(2022)}]{Goulielmakis2022}%
  \BibitemOpen
  \bibfield  {author} {\bibinfo {author} {\bibfnamefont {E.}~\bibnamefont
  {Goulielmakis}}\ and\ \bibinfo {author} {\bibfnamefont {T.}~\bibnamefont
  {Brabec}},\ }\bibfield  {title} {\bibinfo {title} {High harmonic generation
  in condensed matter},\ }\href {https://doi.org/10.1038/s41566-022-00988-y}
  {\bibfield  {journal} {\bibinfo  {journal} {Nature Photonics}\ }\textbf
  {\bibinfo {volume} {16}},\ \bibinfo {pages} {411} (\bibinfo {year}
  {2022})}\BibitemShut {NoStop}%
\bibitem [{\citenamefont {Bauer}\ and\ \citenamefont
  {Hansen}(2018)}]{bauer_high-harmonic_2018}%
  \BibitemOpen
  \bibfield  {author} {\bibinfo {author} {\bibfnamefont {D.}~\bibnamefont
  {Bauer}}\ and\ \bibinfo {author} {\bibfnamefont {K.~K.}\ \bibnamefont
  {Hansen}},\ }\bibfield  {title} {\bibinfo {title} {High-harmonic generation
  in solids with and without topological edge states},\ }\href
  {https://doi.org/10.1103/PhysRevLett.120.177401} {\bibfield  {journal}
  {\bibinfo  {journal} {Phys. Rev. Lett.}\ }\textbf {\bibinfo {volume} {120}},\
  \bibinfo {pages} {177401} (\bibinfo {year} {2018})}\BibitemShut {NoStop}%
\bibitem [{\citenamefont {Dr\"ueke}\ and\ \citenamefont
  {Bauer}(2019)}]{DrueekeBauer2019}%
  \BibitemOpen
  \bibfield  {author} {\bibinfo {author} {\bibfnamefont {H.}~\bibnamefont
  {Dr\"ueke}}\ and\ \bibinfo {author} {\bibfnamefont {D.}~\bibnamefont
  {Bauer}},\ }\bibfield  {title} {\bibinfo {title} {Robustness of topologically
  sensitive harmonic generation in laser-driven linear chains},\ }\href
  {https://doi.org/10.1103/PhysRevA.99.053402} {\bibfield  {journal} {\bibinfo
  {journal} {Phys. Rev. A}\ }\textbf {\bibinfo {volume} {99}},\ \bibinfo
  {pages} {053402} (\bibinfo {year} {2019})}\BibitemShut {NoStop}%
\bibitem [{\citenamefont {J\"ur\ss{}}\ and\ \citenamefont
  {Bauer}(2019)}]{JuerssBauer2019}%
  \BibitemOpen
  \bibfield  {author} {\bibinfo {author} {\bibfnamefont {C.}~\bibnamefont
  {J\"ur\ss{}}}\ and\ \bibinfo {author} {\bibfnamefont {D.}~\bibnamefont
  {Bauer}},\ }\bibfield  {title} {\bibinfo {title} {{High-harmonic generation
  in Su-Schrieffer-Heeger chains}},\ }\href
  {https://doi.org/10.1103/PhysRevB.99.195428} {\bibfield  {journal} {\bibinfo
  {journal} {Phys. Rev. B}\ }\textbf {\bibinfo {volume} {99}},\ \bibinfo
  {pages} {195428} (\bibinfo {year} {2019})}\BibitemShut {NoStop}%
\bibitem [{\citenamefont {Bai}\ \emph {et~al.}(2021)\citenamefont {Bai},
  \citenamefont {Fei}, \citenamefont {Wang}, \citenamefont {Li}, \citenamefont
  {Li}, \citenamefont {Song}, \citenamefont {Li}, \citenamefont {Xu},\ and\
  \citenamefont {Liu}}]{Bai2021}%
  \BibitemOpen
  \bibfield  {author} {\bibinfo {author} {\bibfnamefont {Y.}~\bibnamefont
  {Bai}}, \bibinfo {author} {\bibfnamefont {F.}~\bibnamefont {Fei}}, \bibinfo
  {author} {\bibfnamefont {S.}~\bibnamefont {Wang}}, \bibinfo {author}
  {\bibfnamefont {N.}~\bibnamefont {Li}}, \bibinfo {author} {\bibfnamefont
  {X.}~\bibnamefont {Li}}, \bibinfo {author} {\bibfnamefont {F.}~\bibnamefont
  {Song}}, \bibinfo {author} {\bibfnamefont {R.}~\bibnamefont {Li}}, \bibinfo
  {author} {\bibfnamefont {Z.}~\bibnamefont {Xu}},\ and\ \bibinfo {author}
  {\bibfnamefont {P.}~\bibnamefont {Liu}},\ }\bibfield  {title} {\bibinfo
  {title} {High-harmonic generation from topological surface states},\ }\href
  {https://doi.org/10.1038/s41567-020-01052-8} {\bibfield  {journal} {\bibinfo
  {journal} {Nature Physics}\ }\textbf {\bibinfo {volume} {17}},\ \bibinfo
  {pages} {311} (\bibinfo {year} {2021})}\BibitemShut {NoStop}%
\bibitem [{\citenamefont {Schmid}\ \emph {et~al.}(2021)\citenamefont {Schmid},
  \citenamefont {Weigl}, \citenamefont {Gr{\"o}ssing}, \citenamefont {Junk},
  \citenamefont {Gorini}, \citenamefont {Schlauderer}, \citenamefont {Ito},
  \citenamefont {Meierhofer}, \citenamefont {Hofmann}, \citenamefont
  {Afanasiev}, \citenamefont {Crewse}, \citenamefont {Kokh}, \citenamefont
  {Tereshchenko}, \citenamefont {G{\"u}dde}, \citenamefont {Evers},
  \citenamefont {Wilhelm}, \citenamefont {Richter}, \citenamefont {H{\"o}fer},\
  and\ \citenamefont {Huber}}]{Schmid2021}%
  \BibitemOpen
  \bibfield  {author} {\bibinfo {author} {\bibfnamefont {C.~P.}\ \bibnamefont
  {Schmid}}, \bibinfo {author} {\bibfnamefont {L.}~\bibnamefont {Weigl}},
  \bibinfo {author} {\bibfnamefont {P.}~\bibnamefont {Gr{\"o}ssing}}, \bibinfo
  {author} {\bibfnamefont {V.}~\bibnamefont {Junk}}, \bibinfo {author}
  {\bibfnamefont {C.}~\bibnamefont {Gorini}}, \bibinfo {author} {\bibfnamefont
  {S.}~\bibnamefont {Schlauderer}}, \bibinfo {author} {\bibfnamefont
  {S.}~\bibnamefont {Ito}}, \bibinfo {author} {\bibfnamefont {M.}~\bibnamefont
  {Meierhofer}}, \bibinfo {author} {\bibfnamefont {N.}~\bibnamefont {Hofmann}},
  \bibinfo {author} {\bibfnamefont {D.}~\bibnamefont {Afanasiev}}, \bibinfo
  {author} {\bibfnamefont {J.}~\bibnamefont {Crewse}}, \bibinfo {author}
  {\bibfnamefont {K.~A.}\ \bibnamefont {Kokh}}, \bibinfo {author}
  {\bibfnamefont {O.~E.}\ \bibnamefont {Tereshchenko}}, \bibinfo {author}
  {\bibfnamefont {J.}~\bibnamefont {G{\"u}dde}}, \bibinfo {author}
  {\bibfnamefont {F.}~\bibnamefont {Evers}}, \bibinfo {author} {\bibfnamefont
  {J.}~\bibnamefont {Wilhelm}}, \bibinfo {author} {\bibfnamefont
  {K.}~\bibnamefont {Richter}}, \bibinfo {author} {\bibfnamefont
  {U.}~\bibnamefont {H{\"o}fer}},\ and\ \bibinfo {author} {\bibfnamefont
  {R.}~\bibnamefont {Huber}},\ }\bibfield  {title} {\bibinfo {title} {Tunable
  non-integer high-harmonic generation in a topological insulator},\ }\href
  {https://doi.org/10.1038/s41586-021-03466-7} {\bibfield  {journal} {\bibinfo
  {journal} {Nature}\ }\textbf {\bibinfo {volume} {593}},\ \bibinfo {pages}
  {385} (\bibinfo {year} {2021})}\BibitemShut {NoStop}%
\bibitem [{\citenamefont {J{\"u}r{\ss}}\ and\ \citenamefont
  {Bauer}(2021{\natexlab{a}})}]{Juerss_edge_nanoribbon}%
  \BibitemOpen
  \bibfield  {author} {\bibinfo {author} {\bibfnamefont {C.}~\bibnamefont
  {J{\"u}r{\ss}}}\ and\ \bibinfo {author} {\bibfnamefont {D.}~\bibnamefont
  {Bauer}},\ }\bibfield  {title} {\bibinfo {title} {Edge-state influence on
  high-order harmonic generation in topological nanoribbons},\ }\href
  {https://doi.org/10.1140/epjd/s10053-021-00201-9} {\bibfield  {journal}
  {\bibinfo  {journal} {The European Physical Journal D}\ }\textbf {\bibinfo
  {volume} {75}},\ \bibinfo {pages} {190} (\bibinfo {year}
  {2021}{\natexlab{a}})}\BibitemShut {NoStop}%
\bibitem [{\citenamefont {Hasan}\ and\ \citenamefont
  {Kane}(2010)}]{topinsRevModPhys.82.3045}%
  \BibitemOpen
  \bibfield  {author} {\bibinfo {author} {\bibfnamefont {M.~Z.}\ \bibnamefont
  {Hasan}}\ and\ \bibinfo {author} {\bibfnamefont {C.~L.}\ \bibnamefont
  {Kane}},\ }\bibfield  {title} {\bibinfo {title} {{Colloquium: Topological
  insulators}},\ }\href {https://doi.org/10.1103/RevModPhys.82.3045} {\bibfield
   {journal} {\bibinfo  {journal} {Rev. Mod. Phys.}\ }\textbf {\bibinfo
  {volume} {82}},\ \bibinfo {pages} {3045} (\bibinfo {year}
  {2010})}\BibitemShut {NoStop}%
\bibitem [{\citenamefont {Haldane}(1988)}]{Haldane_1988}%
  \BibitemOpen
  \bibfield  {author} {\bibinfo {author} {\bibfnamefont {F.~D.~M.}\
  \bibnamefont {Haldane}},\ }\bibfield  {title} {\bibinfo {title} {{Model for a
  Quantum Hall Effect without Landau Levels: Condensed-Matter Realization of
  the "Parity Anomaly"}},\ }\href {https://doi.org/10.1103/PhysRevLett.61.2015}
  {\bibfield  {journal} {\bibinfo  {journal} {Phys. Rev. Lett.}\ }\textbf
  {\bibinfo {volume} {61}},\ \bibinfo {pages} {2015} (\bibinfo {year}
  {1988})}\BibitemShut {NoStop}%
\bibitem [{\citenamefont {Kane}\ and\ \citenamefont
  {Mele}(2005)}]{QSHE_Graphene}%
  \BibitemOpen
  \bibfield  {author} {\bibinfo {author} {\bibfnamefont {C.~L.}\ \bibnamefont
  {Kane}}\ and\ \bibinfo {author} {\bibfnamefont {E.~J.}\ \bibnamefont
  {Mele}},\ }\bibfield  {title} {\bibinfo {title} {Quantum spin hall effect in
  graphene},\ }\href {https://doi.org/10.1103/PhysRevLett.95.226801} {\bibfield
   {journal} {\bibinfo  {journal} {Phys. Rev. Lett.}\ }\textbf {\bibinfo
  {volume} {95}},\ \bibinfo {pages} {226801} (\bibinfo {year}
  {2005})}\BibitemShut {NoStop}%
\bibitem [{\citenamefont {Jotzu}\ \emph {et~al.}(2014)\citenamefont {Jotzu},
  \citenamefont {Messer}, \citenamefont {Desbuquois}, \citenamefont {Lebrat},
  \citenamefont {Uehlinger}, \citenamefont {Greif},\ and\ \citenamefont
  {Esslinger}}]{Jotzu2014}%
  \BibitemOpen
  \bibfield  {author} {\bibinfo {author} {\bibfnamefont {G.}~\bibnamefont
  {Jotzu}}, \bibinfo {author} {\bibfnamefont {M.}~\bibnamefont {Messer}},
  \bibinfo {author} {\bibfnamefont {R.}~\bibnamefont {Desbuquois}}, \bibinfo
  {author} {\bibfnamefont {M.}~\bibnamefont {Lebrat}}, \bibinfo {author}
  {\bibfnamefont {T.}~\bibnamefont {Uehlinger}}, \bibinfo {author}
  {\bibfnamefont {D.}~\bibnamefont {Greif}},\ and\ \bibinfo {author}
  {\bibfnamefont {T.}~\bibnamefont {Esslinger}},\ }\bibfield  {title} {\bibinfo
  {title} {{Experimental realization of the topological Haldane model with
  ultracold fermions}},\ }\href {https://doi.org/10.1038/nature13915}
  {\bibfield  {journal} {\bibinfo  {journal} {Nature}\ }\textbf {\bibinfo
  {volume} {515}},\ \bibinfo {pages} {237} (\bibinfo {year}
  {2014})}\BibitemShut {NoStop}%
\bibitem [{\citenamefont {Rechtsman}\ \emph {et~al.}(2013)\citenamefont
  {Rechtsman}, \citenamefont {Zeuner}, \citenamefont {Plotnik}, \citenamefont
  {Lumer}, \citenamefont {Podolsky}, \citenamefont {Dreisow}, \citenamefont
  {Nolte}, \citenamefont {Segev},\ and\ \citenamefont
  {Szameit}}]{Rechtsman2013}%
  \BibitemOpen
  \bibfield  {author} {\bibinfo {author} {\bibfnamefont {M.~C.}\ \bibnamefont
  {Rechtsman}}, \bibinfo {author} {\bibfnamefont {J.~M.}\ \bibnamefont
  {Zeuner}}, \bibinfo {author} {\bibfnamefont {Y.}~\bibnamefont {Plotnik}},
  \bibinfo {author} {\bibfnamefont {Y.}~\bibnamefont {Lumer}}, \bibinfo
  {author} {\bibfnamefont {D.}~\bibnamefont {Podolsky}}, \bibinfo {author}
  {\bibfnamefont {F.}~\bibnamefont {Dreisow}}, \bibinfo {author} {\bibfnamefont
  {S.}~\bibnamefont {Nolte}}, \bibinfo {author} {\bibfnamefont
  {M.}~\bibnamefont {Segev}},\ and\ \bibinfo {author} {\bibfnamefont
  {A.}~\bibnamefont {Szameit}},\ }\bibfield  {title} {\bibinfo {title}
  {{Photonic Floquet topological insulators}},\ }\href
  {https://doi.org/10.1038/nature12066} {\bibfield  {journal} {\bibinfo
  {journal} {Nature}\ }\textbf {\bibinfo {volume} {496}},\ \bibinfo {pages}
  {196} (\bibinfo {year} {2013})}\BibitemShut {NoStop}%
\bibitem [{\citenamefont {Hao}\ \emph {et~al.}(2008)\citenamefont {Hao},
  \citenamefont {Zhang}, \citenamefont {Wang}, \citenamefont {Zhang},\ and\
  \citenamefont {Wang}}]{Hao2008}%
  \BibitemOpen
  \bibfield  {author} {\bibinfo {author} {\bibfnamefont {N.}~\bibnamefont
  {Hao}}, \bibinfo {author} {\bibfnamefont {P.}~\bibnamefont {Zhang}}, \bibinfo
  {author} {\bibfnamefont {Z.}~\bibnamefont {Wang}}, \bibinfo {author}
  {\bibfnamefont {W.}~\bibnamefont {Zhang}},\ and\ \bibinfo {author}
  {\bibfnamefont {Y.}~\bibnamefont {Wang}},\ }\bibfield  {title} {\bibinfo
  {title} {{Topological edge states and quantum Hall effect in the Haldane
  model}},\ }\href {https://doi.org/10.1103/PhysRevB.78.075438} {\bibfield
  {journal} {\bibinfo  {journal} {Phys. Rev. B}\ }\textbf {\bibinfo {volume}
  {78}},\ \bibinfo {pages} {075438} (\bibinfo {year} {2008})}\BibitemShut
  {NoStop}%
\bibitem [{\citenamefont {Yao}\ \emph {et~al.}(2009)\citenamefont {Yao},
  \citenamefont {Yang},\ and\ \citenamefont {Niu}}]{Yao_edgeStatesGraphene}%
  \BibitemOpen
  \bibfield  {author} {\bibinfo {author} {\bibfnamefont {W.}~\bibnamefont
  {Yao}}, \bibinfo {author} {\bibfnamefont {S.~A.}\ \bibnamefont {Yang}},\ and\
  \bibinfo {author} {\bibfnamefont {Q.}~\bibnamefont {Niu}},\ }\bibfield
  {title} {\bibinfo {title} {{Edge States in Graphene: From Gapped Flat-Band to
  Gapless Chiral Modes}},\ }\href
  {https://doi.org/10.1103/PhysRevLett.102.096801} {\bibfield  {journal}
  {\bibinfo  {journal} {Phys. Rev. Lett.}\ }\textbf {\bibinfo {volume} {102}},\
  \bibinfo {pages} {096801} (\bibinfo {year} {2009})}\BibitemShut {NoStop}%
\bibitem [{\citenamefont {Silva}\ \emph {et~al.}(2019)\citenamefont {Silva},
  \citenamefont {Jim{\'e}nez-Gal{\'a}n}, \citenamefont {Amorim}, \citenamefont
  {Smirnova},\ and\ \citenamefont {Ivanov}}]{Silva2019}%
  \BibitemOpen
  \bibfield  {author} {\bibinfo {author} {\bibfnamefont {R.~E.~F.}\
  \bibnamefont {Silva}}, \bibinfo {author} {\bibfnamefont {{\'A}.}~\bibnamefont
  {Jim{\'e}nez-Gal{\'a}n}}, \bibinfo {author} {\bibfnamefont {B.}~\bibnamefont
  {Amorim}}, \bibinfo {author} {\bibfnamefont {O.}~\bibnamefont {Smirnova}},\
  and\ \bibinfo {author} {\bibfnamefont {M.}~\bibnamefont {Ivanov}},\
  }\bibfield  {title} {\bibinfo {title} {Topological strong-field physics on
  sub-laser-cycle timescale},\ }\href
  {https://doi.org/10.1038/s41566-019-0516-1} {\bibfield  {journal} {\bibinfo
  {journal} {Nature Photonics}\ }\textbf {\bibinfo {volume} {13}},\ \bibinfo
  {pages} {849} (\bibinfo {year} {2019})}\BibitemShut {NoStop}%
\bibitem [{\citenamefont {Chac\'on}\ \emph {et~al.}(2020)\citenamefont
  {Chac\'on}, \citenamefont {Kim}, \citenamefont {Zhu}, \citenamefont {Kelly},
  \citenamefont {Dauphin}, \citenamefont {Pisanty}, \citenamefont {Maxwell},
  \citenamefont {Pic\'on}, \citenamefont {Ciappina}, \citenamefont {Kim},
  \citenamefont {Ticknor}, \citenamefont {Saxena},\ and\ \citenamefont
  {Lewenstein}}]{chacon_observing_2018}%
  \BibitemOpen
  \bibfield  {author} {\bibinfo {author} {\bibfnamefont {A.}~\bibnamefont
  {Chac\'on}}, \bibinfo {author} {\bibfnamefont {D.}~\bibnamefont {Kim}},
  \bibinfo {author} {\bibfnamefont {W.}~\bibnamefont {Zhu}}, \bibinfo {author}
  {\bibfnamefont {S.~P.}\ \bibnamefont {Kelly}}, \bibinfo {author}
  {\bibfnamefont {A.}~\bibnamefont {Dauphin}}, \bibinfo {author} {\bibfnamefont
  {E.}~\bibnamefont {Pisanty}}, \bibinfo {author} {\bibfnamefont {A.~S.}\
  \bibnamefont {Maxwell}}, \bibinfo {author} {\bibfnamefont {A.}~\bibnamefont
  {Pic\'on}}, \bibinfo {author} {\bibfnamefont {M.~F.}\ \bibnamefont
  {Ciappina}}, \bibinfo {author} {\bibfnamefont {D.~E.}\ \bibnamefont {Kim}},
  \bibinfo {author} {\bibfnamefont {C.}~\bibnamefont {Ticknor}}, \bibinfo
  {author} {\bibfnamefont {A.}~\bibnamefont {Saxena}},\ and\ \bibinfo {author}
  {\bibfnamefont {M.}~\bibnamefont {Lewenstein}},\ }\bibfield  {title}
  {\bibinfo {title} {{Circular dichroism in higher-order harmonic generation:
  Heralding topological phases and transitions in Chern insulators}},\ }\href
  {https://doi.org/10.1103/PhysRevB.102.134115} {\bibfield  {journal} {\bibinfo
   {journal} {Phys. Rev. B}\ }\textbf {\bibinfo {volume} {102}},\ \bibinfo
  {pages} {134115} (\bibinfo {year} {2020})}\BibitemShut {NoStop}%
\bibitem [{\citenamefont {J\"ur\ss{}}\ and\ \citenamefont
  {Bauer}(2020)}]{juerss2020helicity}%
  \BibitemOpen
  \bibfield  {author} {\bibinfo {author} {\bibfnamefont {C.}~\bibnamefont
  {J\"ur\ss{}}}\ and\ \bibinfo {author} {\bibfnamefont {D.}~\bibnamefont
  {Bauer}},\ }\bibfield  {title} {\bibinfo {title} {{Helicity flip of
  high-order harmonic photons in Haldane nanoribbons}},\ }\href
  {https://doi.org/10.1103/PhysRevA.102.043105} {\bibfield  {journal} {\bibinfo
   {journal} {Phys. Rev. A}\ }\textbf {\bibinfo {volume} {102}},\ \bibinfo
  {pages} {043105} (\bibinfo {year} {2020})}\BibitemShut {NoStop}%
\bibitem [{\citenamefont {Baykusheva}\ \emph
  {et~al.}(2021{\natexlab{a}})\citenamefont {Baykusheva}, \citenamefont
  {Chac\'on}, \citenamefont {Kim}, \citenamefont {Kim}, \citenamefont {Reis},\
  and\ \citenamefont {Ghimire}}]{Ghimire2021_PRA}%
  \BibitemOpen
  \bibfield  {author} {\bibinfo {author} {\bibfnamefont {D.}~\bibnamefont
  {Baykusheva}}, \bibinfo {author} {\bibfnamefont {A.}~\bibnamefont
  {Chac\'on}}, \bibinfo {author} {\bibfnamefont {D.}~\bibnamefont {Kim}},
  \bibinfo {author} {\bibfnamefont {D.~E.}\ \bibnamefont {Kim}}, \bibinfo
  {author} {\bibfnamefont {D.~A.}\ \bibnamefont {Reis}},\ and\ \bibinfo
  {author} {\bibfnamefont {S.}~\bibnamefont {Ghimire}},\ }\bibfield  {title}
  {\bibinfo {title} {Strong-field physics in three-dimensional topological
  insulators},\ }\href {https://doi.org/10.1103/PhysRevA.103.023101} {\bibfield
   {journal} {\bibinfo  {journal} {Phys. Rev. A}\ }\textbf {\bibinfo {volume}
  {103}},\ \bibinfo {pages} {023101} (\bibinfo {year}
  {2021}{\natexlab{a}})}\BibitemShut {NoStop}%
\bibitem [{\citenamefont {Baykusheva}\ \emph
  {et~al.}(2021{\natexlab{b}})\citenamefont {Baykusheva}, \citenamefont
  {Chacón}, \citenamefont {Lu}, \citenamefont {Bailey}, \citenamefont
  {Sobota}, \citenamefont {Soifer}, \citenamefont {Kirchmann}, \citenamefont
  {Rotundu}, \citenamefont {Uher}, \citenamefont {Heinz}, \citenamefont
  {Reis},\ and\ \citenamefont {Ghimire}}]{Ghimire2021_nano}%
  \BibitemOpen
  \bibfield  {author} {\bibinfo {author} {\bibfnamefont {D.}~\bibnamefont
  {Baykusheva}}, \bibinfo {author} {\bibfnamefont {A.}~\bibnamefont {Chacón}},
  \bibinfo {author} {\bibfnamefont {J.}~\bibnamefont {Lu}}, \bibinfo {author}
  {\bibfnamefont {T.~P.}\ \bibnamefont {Bailey}}, \bibinfo {author}
  {\bibfnamefont {J.~A.}\ \bibnamefont {Sobota}}, \bibinfo {author}
  {\bibfnamefont {H.}~\bibnamefont {Soifer}}, \bibinfo {author} {\bibfnamefont
  {P.~S.}\ \bibnamefont {Kirchmann}}, \bibinfo {author} {\bibfnamefont
  {C.}~\bibnamefont {Rotundu}}, \bibinfo {author} {\bibfnamefont
  {C.}~\bibnamefont {Uher}}, \bibinfo {author} {\bibfnamefont {T.~F.}\
  \bibnamefont {Heinz}}, \bibinfo {author} {\bibfnamefont {D.~A.}\ \bibnamefont
  {Reis}},\ and\ \bibinfo {author} {\bibfnamefont {S.}~\bibnamefont
  {Ghimire}},\ }\bibfield  {title} {\bibinfo {title} {{All-Optical Probe of
  Three-Dimensional Topological Insulators Based on High-Harmonic Generation by
  Circularly Polarized Laser Fields}},\ }\href
  {https://doi.org/10.1021/acs.nanolett.1c02145} {\bibfield  {journal}
  {\bibinfo  {journal} {Nano Letters}\ }\textbf {\bibinfo {volume} {21}},\
  \bibinfo {pages} {8970} (\bibinfo {year} {2021}{\natexlab{b}})}\BibitemShut
  {NoStop}%
\bibitem [{\citenamefont {J{\"u}r{\ss}}\ and\ \citenamefont
  {Bauer}(2021{\natexlab{b}})}]{juerss_bauer_ArmZZ}%
  \BibitemOpen
  \bibfield  {author} {\bibinfo {author} {\bibfnamefont {C.}~\bibnamefont
  {J{\"u}r{\ss}}}\ and\ \bibinfo {author} {\bibfnamefont {D.}~\bibnamefont
  {Bauer}},\ }\bibfield  {title} {\bibinfo {title} {High-order harmonic
  generation in hexagonal nanoribbons},\ }\href
  {https://doi.org/10.1140/epjs/s11734-021-00106-z} {\bibfield  {journal}
  {\bibinfo  {journal} {The European Physical Journal Special Topics}\ }\textbf
  {\bibinfo {volume} {230}},\ \bibinfo {pages} {4081} (\bibinfo {year}
  {2021}{\natexlab{b}})}\BibitemShut {NoStop}%
\bibitem [{\citenamefont {Graf}\ and\ \citenamefont {Vogl}(1995)}]{Graf_1995}%
  \BibitemOpen
  \bibfield  {author} {\bibinfo {author} {\bibfnamefont {M.}~\bibnamefont
  {Graf}}\ and\ \bibinfo {author} {\bibfnamefont {P.}~\bibnamefont {Vogl}},\
  }\bibfield  {title} {\bibinfo {title} {Electromagnetic fields and dielectric
  response in empirical tight-binding theory},\ }\href
  {https://doi.org/10.1103/PhysRevB.51.4940} {\bibfield  {journal} {\bibinfo
  {journal} {Phys. Rev. B}\ }\textbf {\bibinfo {volume} {51}},\ \bibinfo
  {pages} {4940} (\bibinfo {year} {1995})}\BibitemShut {NoStop}%
\bibitem [{\citenamefont {Press}\ \emph {et~al.}(2007)\citenamefont {Press},
  \citenamefont {Teukolsky}, \citenamefont {Vetterling},\ and\ \citenamefont
  {Flannery}}]{Numerical_recipies}%
  \BibitemOpen
  \bibfield  {author} {\bibinfo {author} {\bibfnamefont {W.}~\bibnamefont
  {Press}}, \bibinfo {author} {\bibfnamefont {S.}~\bibnamefont {Teukolsky}},
  \bibinfo {author} {\bibfnamefont {W.}~\bibnamefont {Vetterling}},\ and\
  \bibinfo {author} {\bibfnamefont {B.}~\bibnamefont {Flannery}},\ }\href@noop
  {} {\emph {\bibinfo {title} {{Numerical Recipes 3rd Edition: The Art of
  Scientific Computing}}}}\ (\bibinfo  {publisher} {Cambridge University
  Press},\ \bibinfo {year} {2007})\BibitemShut {NoStop}%
\bibitem [{\citenamefont {Kuzemsky}(2011)}]{Review_Transport}%
  \BibitemOpen
  \bibfield  {author} {\bibinfo {author} {\bibfnamefont {A.~L.}\ \bibnamefont
  {Kuzemsky}},\ }\bibfield  {title} {\bibinfo {title} {Electronic transport in
  metallic systems and generalized kinetic equations},\ }\href
  {https://doi.org/10.1142/S0217979211059012} {\bibfield  {journal} {\bibinfo
  {journal} {International Journal of Modern Physics B}\ }\textbf {\bibinfo
  {volume} {25}},\ \bibinfo {pages} {3071} (\bibinfo {year}
  {2011})}\BibitemShut {NoStop}%
\bibitem [{\citenamefont {Moos}\ \emph {et~al.}(2020)\citenamefont {Moos},
  \citenamefont {J\"ur\ss{}},\ and\ \citenamefont {Bauer}}]{Moos2020}%
  \BibitemOpen
  \bibfield  {author} {\bibinfo {author} {\bibfnamefont {D.}~\bibnamefont
  {Moos}}, \bibinfo {author} {\bibfnamefont {C.}~\bibnamefont {J\"ur\ss{}}},\
  and\ \bibinfo {author} {\bibfnamefont {D.}~\bibnamefont {Bauer}},\ }\bibfield
   {title} {\bibinfo {title} {Intense-laser-driven electron dynamics and
  high-order harmonic generation in solids including topological effects},\
  }\href {https://doi.org/10.1103/PhysRevA.102.053112} {\bibfield  {journal}
  {\bibinfo  {journal} {Phys. Rev. A}\ }\textbf {\bibinfo {volume} {102}},\
  \bibinfo {pages} {053112} (\bibinfo {year} {2020})}\BibitemShut {NoStop}%
\bibitem [{\citenamefont {Cooper}\ \emph {et~al.}(2012)\citenamefont {Cooper},
  \citenamefont {D'Anjou}, \citenamefont {Ghattamaneni}, \citenamefont
  {Harack}, \citenamefont {Hilke}, \citenamefont {Horth}, \citenamefont
  {Majlis}, \citenamefont {Massicotte}, \citenamefont {Vandsburger},
  \citenamefont {Whiteway},\ and\ \citenamefont {Yu}}]{Cooper_2012}%
  \BibitemOpen
  \bibfield  {author} {\bibinfo {author} {\bibfnamefont {D.~R.}\ \bibnamefont
  {Cooper}}, \bibinfo {author} {\bibfnamefont {B.}~\bibnamefont {D'Anjou}},
  \bibinfo {author} {\bibfnamefont {N.}~\bibnamefont {Ghattamaneni}}, \bibinfo
  {author} {\bibfnamefont {B.}~\bibnamefont {Harack}}, \bibinfo {author}
  {\bibfnamefont {M.}~\bibnamefont {Hilke}}, \bibinfo {author} {\bibfnamefont
  {A.}~\bibnamefont {Horth}}, \bibinfo {author} {\bibfnamefont
  {N.}~\bibnamefont {Majlis}}, \bibinfo {author} {\bibfnamefont
  {M.}~\bibnamefont {Massicotte}}, \bibinfo {author} {\bibfnamefont
  {L.}~\bibnamefont {Vandsburger}}, \bibinfo {author} {\bibfnamefont
  {E.}~\bibnamefont {Whiteway}},\ and\ \bibinfo {author} {\bibfnamefont
  {V.}~\bibnamefont {Yu}},\ }\bibfield  {title} {\bibinfo {title}
  {{Experimental Review of Graphene}},\ }\href@noop {} {\bibfield  {journal}
  {\bibinfo  {journal} {ISRN Condensed Matter Physics}\ }\textbf {\bibinfo
  {volume} {2012}} (\bibinfo {year} {2012})},\ \bibinfo {note} {article ID
  501686}\BibitemShut {NoStop}%
\bibitem [{\citenamefont {Yu}\ \emph {et~al.}(2019)\citenamefont {Yu},
  \citenamefont {Hansen},\ and\ \citenamefont {Madsen}}]{HHG_doped_solids}%
  \BibitemOpen
  \bibfield  {author} {\bibinfo {author} {\bibfnamefont {C.}~\bibnamefont
  {Yu}}, \bibinfo {author} {\bibfnamefont {K.~K.}\ \bibnamefont {Hansen}},\
  and\ \bibinfo {author} {\bibfnamefont {L.~B.}\ \bibnamefont {Madsen}},\
  }\bibfield  {title} {\bibinfo {title} {Enhanced high-order harmonic
  generation in donor-doped band-gap materials},\ }\href
  {https://doi.org/10.1103/PhysRevA.99.013435} {\bibfield  {journal} {\bibinfo
  {journal} {Phys. Rev. A}\ }\textbf {\bibinfo {volume} {99}},\ \bibinfo
  {pages} {013435} (\bibinfo {year} {2019})}\BibitemShut {NoStop}%
\bibitem [{\citenamefont {Nefedova}\ \emph {et~al.}(2021)\citenamefont
  {Nefedova}, \citenamefont {Fröhlich}, \citenamefont {Navarrete},
  \citenamefont {Tancogne-Dejean}, \citenamefont {Franz}, \citenamefont
  {Hamdou}, \citenamefont {Kaassamani}, \citenamefont {Gauthier}, \citenamefont
  {Nicolas}, \citenamefont {Jargot}, \citenamefont {Hanna}, \citenamefont
  {Georges}, \citenamefont {Ciappina}, \citenamefont {Thumm}, \citenamefont
  {Boutu},\ and\ \citenamefont {Merdji}}]{HHG_Doped_MgO-Cr}%
  \BibitemOpen
  \bibfield  {author} {\bibinfo {author} {\bibfnamefont {V.~E.}\ \bibnamefont
  {Nefedova}}, \bibinfo {author} {\bibfnamefont {S.}~\bibnamefont {Fröhlich}},
  \bibinfo {author} {\bibfnamefont {F.}~\bibnamefont {Navarrete}}, \bibinfo
  {author} {\bibfnamefont {N.}~\bibnamefont {Tancogne-Dejean}}, \bibinfo
  {author} {\bibfnamefont {D.}~\bibnamefont {Franz}}, \bibinfo {author}
  {\bibfnamefont {A.}~\bibnamefont {Hamdou}}, \bibinfo {author} {\bibfnamefont
  {S.}~\bibnamefont {Kaassamani}}, \bibinfo {author} {\bibfnamefont
  {D.}~\bibnamefont {Gauthier}}, \bibinfo {author} {\bibfnamefont
  {R.}~\bibnamefont {Nicolas}}, \bibinfo {author} {\bibfnamefont
  {G.}~\bibnamefont {Jargot}}, \bibinfo {author} {\bibfnamefont
  {M.}~\bibnamefont {Hanna}}, \bibinfo {author} {\bibfnamefont
  {P.}~\bibnamefont {Georges}}, \bibinfo {author} {\bibfnamefont {M.~F.}\
  \bibnamefont {Ciappina}}, \bibinfo {author} {\bibfnamefont {U.}~\bibnamefont
  {Thumm}}, \bibinfo {author} {\bibfnamefont {W.}~\bibnamefont {Boutu}},\ and\
  \bibinfo {author} {\bibfnamefont {H.}~\bibnamefont {Merdji}},\ }\bibfield
  {title} {\bibinfo {title} {Enhanced extreme ultraviolet high-harmonic
  generation from chromium-doped magnesium oxide},\ }\href
  {https://doi.org/10.1063/5.0047421} {\bibfield  {journal} {\bibinfo
  {journal} {Applied Physics Letters}\ }\textbf {\bibinfo {volume} {118}},\
  \bibinfo {pages} {201103} (\bibinfo {year} {2021})}\BibitemShut {NoStop}%
\bibitem [{Sup()}]{Suppl_material}%
  \BibitemOpen
  \href@noop {} {}\bibinfo {note} {See Supplemental Material at [URL will be
  inserted by publisher] for a short comparison between the results of the
  topological and trivial system.}\BibitemShut {Stop}%
\end{thebibliography}%

\end{document}